\documentclass[letter,journal]{IEEEtran}
\usepackage{stfloats}
\usepackage{amssymb}
\usepackage{amsmath}
\usepackage{amsfonts}
\usepackage{eucal}
\usepackage{bbm}
\usepackage{amsfonts}
\usepackage[dvips]{graphicx}
\usepackage{multirow}
\usepackage{algorithm}
\usepackage{algorithmic}
\usepackage{cite}
\usepackage[bookmarks=false]{}
\usepackage{amsthm}
\usepackage{algorithm}
\usepackage{algorithmic}
\usepackage{mathbbol}
\usepackage{amsfonts}
\usepackage{graphicx}
\usepackage{amsmath}
\usepackage{amssymb}
\usepackage{latexsym}
\usepackage{graphicx}
\usepackage{cite}
\usepackage{url}
\usepackage{stfloats}
\usepackage{mathrsfs}
\usepackage{setspace}
\usepackage{booktabs}
\usepackage {latexsym}
\usepackage {graphicx}
\usepackage {cite}
\usepackage {color}
\usepackage {url}
\usepackage {mathrsfs}
\theoremstyle{plain}

\usepackage[figuresright]{rotating}

\usepackage{cite}
\usepackage{setspace}
\usepackage{comment}
\usepackage{CJKutf8}
\usepackage{caption}
\usepackage{makecell}
\usepackage{graphicx}
 \usepackage{threeparttable}

\usepackage[tight,footnotesize]{subfigure}
\usepackage{colortbl}
\usepackage{float}
\usepackage{lipsum}



\begin{document}
\begin{CJK}{UTF8}{gbsn}

\title{\huge Positioning Using Wireless Networks: Applications, Recent Progress and Future Challenges

\author{\IEEEauthorblockN{Yang Yang, \textit{Member, IEEE}, Mingzhe Chen,  \textit{Member, IEEE}, Yufei Blankenship, Jemin Lee, \textit{Member, IEEE}, Zabih Ghassemlooy, \textit{Senior Member}, \textit{IEEE}, Julian Cheng \textit{Fellow, IEEE}, and Shiwen Mao \textit{Fellow, IEEE}}

}

}

\maketitle

\renewcommand{\thefootnote}{\fnsymbol{footnote}}

\footnotetext{Yang Yang is with the School of Information and Communication Engineering, Beijing University of Posts and Telecommunications, Beijing, China (e-mail: yangyang01@bupt.edu.cn).

M. Chen is with the Department of Electrical and Computer Engineering and Institute for Data Science and Computing, University of Miami, Coral Gables, FL, USA (e-mail: mingzhe.chen@miami.edu).

Yufei Blankenship is with the Ericsson, Schaumburg, Illinois, USA (e-mail:yufei.blankenship@ericsson.com)

Jemin Lee is with School of Electrical and Electronic Engineering, Yonsei University, Seoul, Korea (e-mail: jemin.lee@yonsei.ac.kr)

Zabih Ghassemlooy is with the Optical Communications Research Group,
Faculty of Engineering and Environment, Northumbria University, NE1 8ST
Newcastle upon Tyne, U.K. (e-mail: z.ghassemlooy@northumbria.ac.uk)

Julian Cheng is with the School of Engineering, The University of British
Columbia, Kelowna, BC, Canada. (julian.cheng@ubc.ca)

Shiwen Mao is the Wireless Engineering Research and Education Center, Auburn University, Auburn, AL, USA. (smao@ieee.org)
}

\begin{abstract}
Positioning has recently received considerable attention as a key enabler in emerging applications such as extended reality, unmanned aerial vehicles and smart environments. These applications require both data communication and high-precision positioning, and thus they are particularly well-suited to be offered in wireless networks (WNs). The purpose of this paper is to provide a comprehensive overview of existing works and new trends in the field of positioning techniques from both the academic and industrial perspectives. The paper provides a comprehensive overview of positioning in WNs, covering the background, applications, measurements, state-of-the-art technologies and future challenges. The paper outlines the applications of positioning from the perspectives of public facilities, enterprises and individual users. We investigate the key performance indicators and measurements of positioning systems, followed by the review of the key enabler techniques such as artificial intelligence/large models and adaptive systems. Next, we discuss a number of typical wireless positioning technologies. We extend our overview beyond the academic progress, to include the standardization efforts, and finally, we provide insight into the challenges that remain. The comprehensive overview of exisitng efforts and new trends in the field of positioning from both the academic and industrial communities would be a useful reference to researchers in the field. 
\end{abstract}


\section{Introduction}
\subsection{Background and Motivation}

High-precision positioning has attracted increasing attention in recent years. In emerging applications such as extended reality (XR), unmanned aerial vehicles (UAVs), and smart environments, positioning plays a key role in accurately mapping a real-world environment to a digital world \cite{trevlakis2023localization}. In future 6G networks, it is envisioned that positioning will become a key function, which can improve the performance of communication, computing, and control \cite{saad2019vision}. Therefore, several positioning technologies (PTs) have been proposed in order to enhance the performance of wireless communication networks and satisfy the demanding requirements of emerging applications.

Global navigation satellite system (GNSS) is a prominent PT for positioning. However, using GNSS for positioning faces several key challenges: (i) meeting the stringent real-time requirements of certain emergent applications due to the significant distance of satellites from the earth; and (ii) high attenuation and reduced reliability in indoor environments due to weak satellite signals and obstructed by roofs, walls, and other solid structures. In order to address these challenges, PTs based on wireless networks (WNs), such as cellular networks, WiFi, Bluetooth, and visible light positioning (VLP), are of great value for indoor environments.

Positioning using WNs offers several advantages over GNSS systems including (i) lower latency, due to the shorter signal propagation time of WNs compared to that of satellites; (ii) improved coverage in indoor environments with a higher level of reliability; (iii) reuse of existing WN infrastructure, which makes the solution more cost-effective; and (iv) enhanced positioning capabilities by using emerging technologies such as artificial intelligence (AI), large foundation models and reconfigurable intelligent surfaces (RIS). Therefore, In consequence, there is an ongoing interaction between new positioning needs and emerging technologies, which necessitate a further comprehensive review of the existing PTs requirements.

\subsection{The Evolution Of PT Over WNs}

Cellular networks are one of the most representative types of WNs, where the early generations were mainly designed for communication services. Historically, positioning has been considered as a byproduct of communications in 1G to 4G, resulting in a limited level of positioning accuracy (PA). For instance, during the 1970s, researchers attempted to locate vehicles using 1G cellular networks based on signal strength, since communication processes such as cell site selection would benefit from knowing the location of the vehicle \cite{ott1977vehicle}. During the 2G era, as the standard lacked a built-in location mechanism, global system for mobile communications (GSM) positioning capabilities were confined to using training or synchronization signals for computing ranging measurements. Release 4 of 3G, as described in TS 22.071, introduced location services with a horizontal location accuracy ranging from 25 to 200 m \cite{del2017survey}. As a result of limited advancements in positioning in 4G networks, it has been demonstrated to achieve a 50 m horizontal location accuracy as required by the enhanced 911 (e911) location requirements defined by the Federal Communications Commission (FCC) with long-term evolution (LTE) location methods \cite{del2017survey}.

With 5G, location information has become increasingly important, which offers reduced latency and enhanced scalability and robustness. Meanwhile, due to the scarcity of wireless spectrum, technologies using millimeter waves (mmWave) with short wavelengths and large bandwidths are being exploited to provide more sensitive signal measurements and better spatial resolution, resulting in higher PA. The next generation wireless network (i.e., 6G and beyond), will introduce terahertz (THz) and optical (both visible and infrared) bands as enabler technologies with improved PA for both indoor and outdoor environments. 

In 2004 \cite{Horikawa2004Pervasive}, a VLP system based on visible light communication (VLC) was proposed for the first time by Horikawa \emph{et al.} It has since been extensively reported that indoor VLP systems based on light emitting diode (LED) lights with line-Of-sight (LOS) propagation paths and limited multipath interference can achieve centimeter-level PA \cite{wu2021indoor,zhu2024survey,8A}. Furthermore, a number of short-to-medium-range wireless technologies such as WiFi, Bluetooth, radio-frequency identification (RFID), and ultra-wideband (UWB), are also available, which are indispensable for future positioning applications.

\subsection{Relevant Works}
Considering that PTs have received much attention over the past century, there are quite a few articles that provide an overview of this interesting and important topic. In particular, most of the existing surveys are specific to a certain technology \cite{zafari2019survey}, such as cellular networks \cite{del2017survey,trevlakis2023localization}, mmWave \cite{shastri2022review}, THz \cite{chen2022tutorial} and VLP \cite{zhuang2018survey,zhu2024survey,8A}, or certain applications domain such as Internet-of-Things (IoT) [12]. Note that there are already some high-quality, generic survey papers reported in the literature \cite{liu2007survey,gu2009survey,yassin2016recent,zafari2019survey,yang2021survey,trevlakis2023localization}. Specifically, Liu \emph{et al.} \cite{liu2007survey} provided an overview of the wireless indoor positioning systems (IPSs) and their performance metrics with a special emphasis on fingerprinting algorithms. Gu \emph{et al.} \cite{gu2009survey} provided a comprehensive survey of numerous IPSs including commercial products, research-oriented solutions, and evaluation criteria. Yassin \emph{et al.} \cite{yassin2016recent} investigated the theoretical aspects and applications of an IPS. Zafari \emph{et al.} \cite{zafari2019survey} presented a detailed description of different IPSs and technologies, whereas Yang \emph{et al.} \cite{yang2021survey} presented key performance metrics, as well as machine learning (ML)-based and filter-based methods adopted in IPSs. 


Compared with the existing survey papers on indoor positioning\cite{del2017survey,shastri2022review,zhuang2018survey,chen2022tutorial,liu2007survey,gu2009survey,yassin2016recent,zafari2019survey,yang2021survey,8A}, this comprehensive survey paper offers the following key features: (i)  reviewing a number of the latest enabling techniques including ML, large models, RIS, adaptive systems and soft-defined networks (SDN) for positioning; (ii) introducing the standardization progress of positioning in various WNs technologies; (iii) providing a comprehensive evaluation criterion for wireless positioning systems (PSs); and (iv) considering the fusion of different positioning technologies. The organization of this paper is summarized as follows.
\begin{itemize}
  \item In Section II, we discuss a series of possible applications of positioning, and highlight their applications in public utilities, enterprise, and individual users. We summarize the key motivations for PSs by analyzing the needs of these emerging positioning applications.
  \item In Section III, we summarize the key performance indicators (KPIs) of a PS. We highlight privacy and security aspects in KPIs. We also summarize the key measurements used for positioning in this section.
  \item  In Section IV, we introduce the advanced techniques used for positioning including ML, adaptive systems, RIS, and SDN. Especially, we introduce the possible application of large models for indoor positioning.
  \item In Section V, we present different wireless technologies for positioning. We primarily discuss cellular networks, WiFi, Bluetooth, RFID, mmWave, UWB, THz, and visible light. We also discuss the advantages and challenges of each technology.
 \item In Section VI, we summarize the challenges and future research directions of positioning.
 \item In Section VII, we conclude this paper.
\end{itemize}

\section{Applications}
It is becoming increasingly common to use positioning in a range of applications. It is reported that the market size for indoor positioning was approximately \$10.9 billion in 2023, and it is expected to reach \$29.8 billion by the end of 2028, with a compound annual growth rate of 22.3\% during the forecast period \cite{marketsandmarkets2023indoor}. Here, we mainly categorize positioning applications into three types: public provision, enterprise and individual users.
\begin{figure*}[t]
    \centering
    \includegraphics[width=1.0\textwidth]{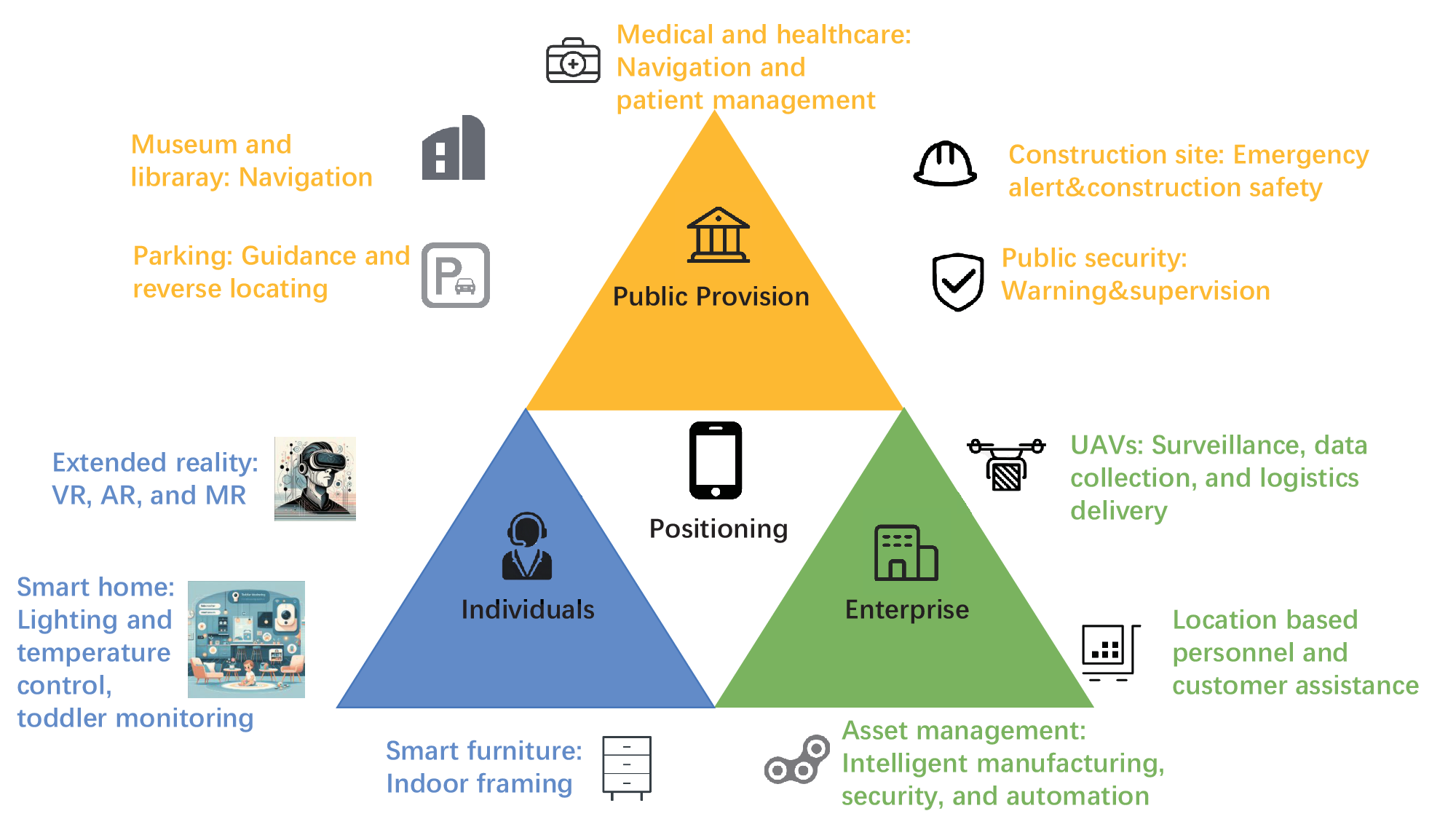}
    \caption{The applications of wireless positioning technologies.}
    \captionsetup{justification = centerlast}
    \label{fig:applications}
\end{figure*}

\subsection{Public Provision}
In public places such as airports, museums and hospitals, indoor positioning enables precise navigation and location-based services, enhancing the visitor experience and maximizing operational efficiency. Public spaces can be made more accessible and manageable using pathfinding, asset tracking, emergency response coordination, and personalized information delivery. 

\emph{Context Aware Location Based User Assistance}: For instance, PSs can enhance visitor experiences by providing location-based services, such as guided tours, information on books, and interactive content directly to visitor’s smartphones or via AR devices. These systems leverage technologies like Wi-Fi, Bluetooth beacons, and RFID to determine the visitor’s location in indoor environments and deliver relevant content accordingly. For example, Huang \emph{et al.} \cite{huang2016get} developed a NO Donkey E-learning system addressing challenges related to spatial and learning domain unawareness and navigation within a library. Similarly, there are other applications such as museums, airports, underground parking, and tourism services, among others, that can benefit from positioning and navigation services.

\emph{Medical and Healthcare}: In the medical and healthcare domain, PSs are also useful to enhance operational efficiency, patient care, and safety. With this capability, not only is asset utilization, tracking, and management optimized and the time spent looking for staff and equipment reduced, but emergency response times are also improved. Furthermore, IPSs enable hospitals to monitor patient movements, ensuring that patients who require special care do not wait for too long and/or end up in restricted areas. Additionally, these systems facilitate navigation within complex hospital buildings, helping patients and visitors to locate departments, wards, and amenities readily. Luschi \emph{et al.} \cite{luschi2022designing} adopted a hybrid mobile application architecture to deploy multiple platforms. It demonstrated that the proposed indoor positioning and navigation system within healthcare facilities can efficiently improve the navigational experience for staff, patients, and visitors.

\emph{Public Security}: A public security application requires the rapid and precise determination of the location of individuals in need of emergency services in order to dispatch police, firefighters, and medical staff to the exact location of the incident as quickly as possible. The positioning delay, robustness, and accuracy are vital in saving lives and reducing the time it takes to provide assistance. Moder \emph{et al.} \cite{moder2018indoor} discussed the use of IPSs in public transport environments to enhance public safety and accessibility for visually impaired people. This study demonstrated the potential of indoor positioning technologies to improve the autonomy and safety of vulnerable populations in complex indoor environments.

\begin{table*}
\renewcommand{\arraystretch}{2}
\centering
\caption{Localization Requirements of Different Applications}
\label{tab233}
\begin{tabular}{|c|l|l|l|l|l|}
\hline
 & \multicolumn{2}{c|}{Application} & \multicolumn{2}{c|}{Requirement} \\
\hline
\multirow{3}{*}{Public provision} & \multicolumn{2}{l|}{\makecell[l]{Context aware location-\\based user assistance}} & \multicolumn{2}{l|}{\makecell[l]{$\bullet$ Meter-level accuracy\\
$\bullet$ Low energy consumption and low cost \cite{table1_6}} 
} \\
\cline{2-5}
 & \multicolumn{2}{l|}{Medical and healthcare} & \multicolumn{2}{l|}{\makecell[l]{$\bullet$ At least meter-level accuracy\\
$\bullet$ High reliability and robustness\\
$\bullet$ Low latency\\
$\bullet$ Privacy information protection \cite{table1_5}}
} \\
\cline{2-5}
 & \multicolumn{2}{l|}{Public security} & \multicolumn{2}{l|}{\makecell[l]{$\bullet$ At least meter-level accuracy\\
$\bullet$ Low latency\\
$\bullet$ High reliability and robustness \cite{moder2018indoor} }
} \\
\hline
\multirow{3}{*}{Enterprise} & \multicolumn{2}{l|}{UAVs} & \multicolumn{2}{l|}{\makecell[l]{$\bullet$ Centimeter-level accuracy\\
$\bullet$ Wide coverage\\
$\bullet$ High mobility tracking \cite{table1_1} and low latency}
} \\
\cline{2-5}
 & \multicolumn{2}{l|}{\makecell[l]{Location based personnel\\and customer assistance}} & \multicolumn{2}{l|}{\makecell[l]{$\bullet$ Meter-level accuracy\\
$\bullet$ Low latency\\
$\bullet$ Wide coverage}
} \\
\cline{2-5}
 & \multicolumn{2}{l|}{\makecell[l]{Asset tracking and management}} & \multicolumn{2}{l|}{\makecell[l]{$\bullet$ At least submeter-level accuracy\\
$\bullet$ Cooperative localization among massive IoT devices \cite{table1_1}\\
$\bullet$ High reliability\\
$\bullet$ Low latency}
} \\
\hline
\multirow{3}{*}{Individuals} & \multicolumn{2}{l|}{Extended reality (XR)} & \multicolumn{2}{l|}{\makecell[l]{$\bullet$ Centimeter-level accuracy (i.e. 1-10 cm)\\
$\bullet$ Very low latency (less than 20 ms) \cite{table1_2} }
} \\
\cline{2-5}
 & \multicolumn{2}{l|}{Smart life} & \multicolumn{2}{l|}{\makecell[l]{$\bullet$ Submeter-level accuracy\\
$\bullet$ NLOS-based localization \cite{table1_3}\\
$\bullet$ Low energy consumption\\
$\bullet$ Privacy information classification and protection \cite{table1_4}}
} \\
\hline
\end{tabular}
\end{table*}

\subsection{Enterprise}
A PS can also be used to improve the operational efficiency of enterprise by providing precise location tracking of staff, UAVs, equipment, assets, and customers.

\emph{UAVs}:
Nowadays, UAVs play an increasingly important role in enhancing the efficiency and accuracy of task execution, offering innovative and safe solutions for data collection, monitoring, and logistics delivery, especially in inaccessible or hazardous environments. For the application of UAVs, the positioning technology is, therefore, of parament importance. For instance, in logistics, UAVs use positioning to streamline delivery routes, demonstrating their pivotal role in automating and optimizing UAV operations across various sectors, from smart manufacturing plants to disaster assessment and beyond \cite{orgeira2020methodology}. Moreover, UAVs themselves can also provide high-precision positioning services. For instance, Wang \emph{et al.} \cite{wang2021toward} proposed a UAV-based PS to provide highly reliable positioning services for people in mountainous environments, where conventional wireless PSs offered limited services. Due to their unique ability to navigate to locations where both the signal propagation conditions and geometric configurations are optimal for positioning, UAVs can outperform conventional ground-based wireless technologies. With the advancement of positioning technology, UAVs are expected to find more applications such as manufacturing, farming, environmental monitoring, and warehouses, among others.

\emph{Location Based Personnel and Customer Assistance}:  
Positioning can optimize paths and tasks based on the employees’ locations and deliver targeted advertisement to users based on their locations, therefore enhancing enterprise efficiency. It can also enhance safety through location-based alerts. Moreover, high-precision positioning allows enterprises to obtain accurate information about their users, thereby increasing their revenue. For example, online advertising is a valuable revenue stream for providers, with location-based advertising emerging as an effective means of enhancing the effectiveness of online advertising. Cheng \emph{et al.} \cite{cheng2020maximizing} investigated a framework to maximize the effectiveness of mobile advertising. The authors concluded that both service providers and customers can benefit from location information.

\emph{Asset Tracking and Management}: Positioning allows enterprises to monitor the location of equipment and assets in real-time, thus reducing inventory tracking and management time and resources. Real-time tracking of goods, for example, ensures transparency from warehouse storage to delivery, enhancing the reliability of supply chains in logistics and supply chain management. In the context of smart factories, indoor positioning is instrumental in optimizing operational efficiency and safety. As a result, it facilitates automated inventory management, enhanced workflow optimization, and the prevention of accidents by ensuring workers do not enter hazardous areas without the proper clearances. By leveraging indoor positioning, factories can achieve higher levels of automation, improve resource allocation, and enhance the overall safety and productivity of their operations \cite{barbieri2021uwb}.

\subsection{Individuals} 
In the online to offline (O2O) ecosystem, PSs play a crucial role in bridging the gap between digital platforms and physical stores. In contrast to the previous subsections that focused on small enterprises and public facilities, this section is primarily concerned with personal services.

\emph{Extended Reality}: 
Positioning is a cornerstone of XR encompassing virtual reality (VR), augmented reality (AR), and mixed reality (MR). For example, in AR applications, PSs enable the overlay of digital content over the real world in a way that seamlessly interacts with the user’s environment. In navigation aids, educational tools, and gaming the alignment of virtual objects with the physical world enhances the user’s sense of presence and engagement \cite{wang2023scene}. For room-scale experiences in VR, location tracking is essential, as it allows users to move freely within a virtual environment that is similar to their physical environment. As a result of this capability, users are not only able to interact more effectively within the virtual domain, but they are also protected from collisions with real-world objects.

\emph{Smart Life}: Positioning is essential in realizing the vision of a smart life, where digital and physical worlds converge in order to enhance the quality of life. In smart homes/offices, location-based technology enables automation systems to adjust lighting, temperature, and security settings in accordance with the residents’  presence or absence, thereby creating a more comfortable and energy-efficient living environment. For personal health, wearable devices use location tracking to monitor physical activities and provide personalized service. For example, A robust PS, for example, can provide rapid location and rescue services for an elderly person who has fallen or assist blind people in navigating within buildings. In the field of transportation, location services can provide real-time navigation, traffic updates and customized personalized travel recommendations, thereby streamlining commutes and reducing congestion.

\subsection{Motivations} 
A summary of the requirements of typical positioning applications is given in Table \ref{tab233}. As can be observed from Table \ref{tab233}, there are different key indicators for different applications. For example, XR typically requires a delay to be within 20 ms and centimeter-level PA. The intelligent adjustment application, on the other hand, does not require such high positioning precision and low latency, but it does require low energy consumption in order to provide long-term service, since many sensors are powered by batteries. As medical and healthcare applications contain life-critical services, the security of patients' data is highly important. Therefore, it is necessary to comprehensively analyze KPIs, techniques, and technologies of PSs, which will be discussed in detail in the following sections.

\section{KPIs and Measurements} 

\subsection{KPIs} 

The primary objective of IPSs is to achieve high levels of PA. There are, however, certain applications that require additional metrics due to their specific features, thus the need for KPIs. An overview of KPIs (accuracy, energy efficiency, availability, cost, latency, scalability, robustness, and security)  for IPSs is provided in this subsection.

\emph{Accuracy}: PSs rely heavily on accuracy, which measures the degree to which the estimated location corresponds to the actual position. It is typically quantified in terms of root mean square error (RMSE) or cumulative distribution function (CDF) of measurements with an error below a specified threshold. There are different levels of PA, that are capable of meeting various business functions, which require a tailored analysis aligned with a specific application scenario. The implementation of location-based store recommendation services, for example, does not require highly accurate location information. Note, a high degree of PA may result in additional costs. An emergent application such as indoor AR navigation, however, will benefit from the higher accuracy of the location information, thus improving the experience for users.

\emph{Energy efficiency}: A crucial performance indicator is the energy efficiency of PSs. This is because a PS that consumes high amount of energy may lead to rapid battery drainage on user devices, thus limiting its application and marketability. As a result, a well-designed PS should be energy efficient, which meets the energy requirement of next-generation wireless networks (i.e., 6G). Note, several factors influence energy efficiency, including transmit power, algorithm complexity, hardware design, etc., so a trade-off must be made between them to achieve the best energy efficiency \cite{zafari2019survey,guo2019survey}. 

\emph{Availability}: Devices as well as services are affected by availability issues. In the former, users can access the PS on their own devices without the need for specialized terminal equipment. For example, WiFi and Bluetooth are widely used technologies that are available on almost all mobile devices. For the latter, availability refers to the continuity and stability of the PS’s services. PSs with good availability should consistently provide accurate, reliable, and real-time positioning services, ensuring users will always receive reliable results regardless of the circumstances.

\emph{Cost}: Cost is an important aspect of the design and application of PSs, which necessitates a thorough consideration of the costs throughout the development process. The PS's cost is influenced by a variety of factors, including hardware expenditures, time investment, human resources, as well as maintenance and expansion of the system. In order to minimize overall cost, an ideal PS should reduce the need for additional infrastructure and avoid relying on high-end user equipment or systems that are difficult to deploy widely.

\emph{Latency}: The term “latency” refers to the amount of time that elapses between sending a request and receiving the corresponding location results. Latency can have a significant impact on the user experience in many real-time applications and can even be life-critical in some circumstances. For example in intelligent transportation systems, the delay in positioning may prevent vehicles from avoiding obstacles or adjusting direction in time, thereby increasing the chance of accidents.

\emph{Scalability}: Scalability refers to a system’s ability to expand geographically and to deal with the increasing number of devices. As the number of users or devices that rely on the PS increases, a system with good scalability should maintain a stable performance and accuracy. As the demand increases. scalability also implies that the system can effectively manage its resources, such as bandwidth and power consumption, in an effective manner. In addition, as technology develops and advances, a scalable PS should also be able to integrate new technologies and standards.  

\emph{Robustness}: 
The robustness of PSs refers to their ability to withstand disturbances and signal losses that may impair their functionalities. In practice, the positioning environment is complex with different situations, such as extreme weather conditions, obstructions, noise and interference, etc. PSs should adapt to different environmental conditions and provide accurate positioning service even in harsh conditions that can affect the transmission of signals.

\emph{Security and Privacy}: Although security has become an increasingly important topic in communications, it is rarely considered a significant indicator in positioning. However, security and privacy issues in PSs are equally important and may compromise other performance metrics such as accuracy and latency as well \cite{sartayeva2023survey}. Security and privacy issues include confidentiality, integrity, authenticity, and other issues such as the draining of resources. Note that different PSs have different safety issues. For example, for beacon-based PSs, the replacement and replaying of beacons can result in inaccurate positioning data. A jamming attack is another example, in which externally introduced noise disrupts the wireless communication channel between the beacon and the receiver. Therefore, security and privacy are also important performance indicators, especially for future intelligent applications that contain a significant amount of personal information and life-critical applications that may lead to serious consequences.

\subsection {Measurement} 
\emph{TOA}: A time of arrival (TOA) or time of flight (TOF)-based distance estimation is based on the propagation time of the signal from the transmitter to the receiver. Two common methods are employed to obtain TOA information. The first method involves estimating the round-trip time (RTT) by including timestamps in the transmission and reception times of the signal. Alternatively, if the system is synchronized, TOA can be directly inferred from the signal, and the time resolution is primarily dependent on the bandwidth of the signal.

\begin{figure}
    \centering
    \includegraphics{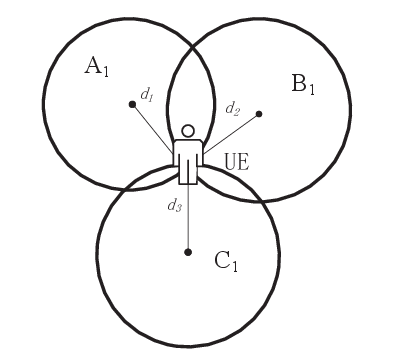}
    \caption{TOA positioning schematic diagram.}
    \captionsetup{justification = centerlast}
    \label{fig:toa}
\end{figure}

ToA-based two-dimensional (2D) positioning algorithms require at least three non-coplanar access points (AP) or anchors, to ensure unique positioning results \cite{buehrer2011fundamentals}, see Fig. \ref{fig:toa}. Assume that the transmitter sends a signal at time 0, and the $i$-th AP receives the signal at time $t_i$. The distance between the transmitter and $i$-th AP can be calculated as $d_i=c \cdot t_i$, where $c=3\times10^{8}$ m/s is the propagation speed. The distances between the three APs and the transmitter are $d_1$, $d_2$, and $d_3$, respectively. Assuming that the AP location is the center and the measured distance is the radius, the target can be located by drawing three circles intersecting at one point. The least square method can be used to calculate the approximate position of the target \cite{guvenc2007analysis}. Alternatively, a three-dimensional (3D) PS requires a minimum of four APs. Khalaf-Allah \emph{et al.} \cite{khalaf2021novel} proposed a solution for the three-anchor ToA-based 3D PS without the need for an initial position guess in order to reduce the hardware deployment costs.

TOA directly measures the arrival time of the signal and can filter out the multipath effects, thereby improving the PA. This technique, however, has the drawback of requiring highly accurate time synchronization between the transmitter and the receiver. A synchronization error of one nanosecond results in a positioning error of 0.3 m \cite{kolodziej2017local}.
The process of achieving synchronization among all units is often challenging and costly, and there are some solutions for PSs using the TOA algorithm when synchronization is imperfect \cite{huang2013efficient}\cite{shi2021moving}. For TOA-based PSs \cite{wang2013position}, position estimation accuracy typically falls within the range of millimeters or centimeters under perfect synchronization between the transmitter and receiver.

\emph{Time difference of arrival (TDOA)}: 
To relieve the strict synchronization requirement of TOA, TDOA is proposed. TDOA determines the transmitter position by measuring the difference in signal arrival time, TOAD is able to determine the transmitter’s position, and, therefore, it only requires strict synchronization between APs or receivers, which is easier to implement. Here, TDOA can either refer to the TDOA of multiple nodes or the TDOA of multiple signals \cite{ferreira2017localization}. In approaches based on TDOA of multiple nodes, several receivers are placed at different locations and kept synchronized in time. Periodically, the transmitter transmits signals and the receivers record the time at which they are received. The time difference between signal arrivals is then calculated. For approaches based on TDOA of multiple signals, the transmitter transmits two different types of signals with different propagation speeds \cite{zhang2010localization}, and the distance between these two devices can be determined by measuring the difference in time between the arrival of these two types of signals \cite{kulaib2011overview}.

\begin{figure}
    \centering
    \includegraphics{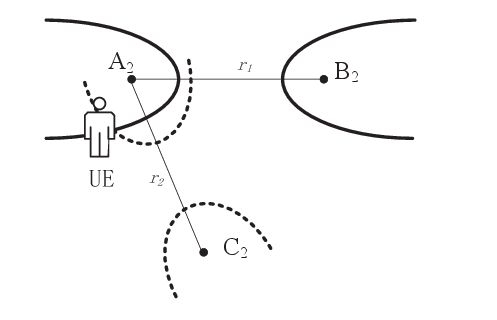}
    \caption{TDOA positioning schematic diagram.}
    \captionsetup{justification = centerlast}
    \label{fig:tdoa}
\end{figure}

The TDOA with multiple nodes requires at least three synchronized APs to locate the transmitter, see Fig. \ref{fig:tdoa}. The technique measures the time difference $t_{i,j}$ between a pair of APs, i.e., AP $i$ and $j$. The distance difference between a pair of APs is defined as $L=c \cdot t_{i,j}$. Using AP $A_2$ and AP $B_2$ as an example, a hyperbola can be obtained by combining the distance difference between the two APs. Similarly, using APs $A_2$ and $C_2$, another hyperbola can be obtained, and the intersection point of the two hyperbolas is the position of the user equipment (UEs)\cite{liu2007survey}. 
The hyperbola can be expressed as:
\begin{equation}\begin{aligned}
L&=\sqrt{\left(x_i-x\right)^2+\left(y_i-y\right)^2+\left(z_i-z\right)^2}\\&-\sqrt{\left(x_j-x\right)^2+\left(y_j-y\right)^2+\left(z_j-z\right)^2},
\end{aligned}\end{equation}
where ($x_i$, $y_i$, $z_i$) is the coordinate of $i$-th AP and ($x$, $y$, $z$) is the coordinate of the transmitter.  The system of hyperbola equations can be solved either through linear regression or by linearizing the equation using Taylor series expansion\cite{jin2018robust}.

In general, the TDOA eliminates the need for synchronization between the transmitter and receiver, which simplifies the PS design and enhances its scalability. However, the PA of TDOA is susceptible to environmental factors such as multipath effects, noise, and non Line-Of-Sight (NLOS).

\begin{figure}
    \centering
    \includegraphics[width=0.5\textwidth]{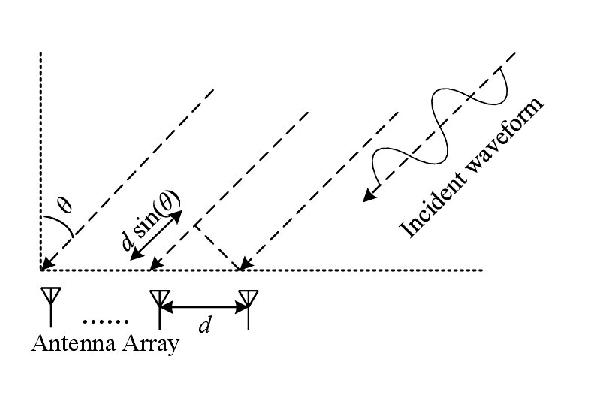}
    \caption{AOA positioning schematic diagram.}
    \captionsetup{justification = centerlast}
    \label{fig:aoa}
\end{figure}

\emph{AOA}: The angle of arrival (AoA) technique estimates the angle of the transmitter to the receiver by equipping the receiver with an antenna array. As shown in Fig. \ref{fig:aoa}, a multi-antenna array produces a time difference in the reception of signals arriving from different angles, which corresponds to the arrival angle of the signals. The AOA algorithm requires at least two APs with known positions. Starting from the AP, the ray formed will pass through the target, which is located at the intersection of the two rays.

AOA-based PS does not require time synchronization and offers higher flexibility compared to TOA or TDOA-based systems, as it only requires the deployment of two APs equipped with antenna arrays. AoA offers accurate estimates, particularly in scenarios where the distance between the transmitter and receiver is relatively short \cite{farahsari2022survey}. For example, in Bluetooth 5.1 standard, the application of AoA greatly enhances the PA. Zhao \emph{et al}. \cite{zhao2022aoa} implemented an AOA IPS based on Bluetooth 5.1 to realize the asset positioning in the warehouse, which has the advantages of simplicity, low installation costs, and sub-meter PA.

The practical implementation of the AOA technology faces several challenges. In the absence of AOA not combined with distance information, only a relative coordinate system can be used for estimating a position. For accurate positioning, hybrid algorithms integrating AOA with other PS, such as TOA or TDOA, have been proposed \cite{blanco2019performance,geng2021experimental,panwar2022optimal}.  Moreover, blockage and multipath propagation may result in inaccurate estimation of AOA.

\emph{POA}: 
The phase of arrival (POA) utilizes the phase of the carrier signal to estimate the distance between the transmitter and the receiver. POA measures the phase of the signal at the receiver, which is modulated with different frequencies and has the same initial phase at the transmitter. By calculating the phase difference, the distance between the two can be obtained. The POA measurement can be combined with ToA, TDoA, or RSSI to improve the accuracy and performance of the PSs. Since distance information is related to the signal phase, POA has relatively lower requirements for signal synchronization, thereby avoiding the impact of time synchronization inaccuracies on positioning results. POA-based methods have the disadvantage of requiring LOS propagation paths to achieve high precision positioning, which is challenging to realize in real-world situations \cite{zafari2019survey}.

\emph{RSS}: 
Received signal strength (RSS) is one of the most popular measurements in IPS due to its simplicity and low costs. Here, RSS typically refers to the absolute measure of the signal strength in dBm or mW at the receiver. A concept closely related to RSS is RSS indicator (RSSI), which is a relative measure of RSS in arbitrary units. Often, RSSI values are often mapped to a scale defined by the hardware manufacturer, which makes is easier to interpret than RSS values. For instance, Atheros WiFi chipset utilizes an RSSI range between 0 and 60; Cisco, on the other hand, uses a broader RSSI range of 0 to 100\cite{zafari2019survey}.

As we know, RSS attenuates with the transmission distance. Therefore, given the channel model, the distance between the transmitter and the receiver can be determined. Here, the channel model varies depending on the types of transmission signals. For instance, for visible light signals, the deterministic Lambertian channel model is typically employed to calculate the distance between the transmitter and the receiver \cite{bai2020received}, while for terahertz (THz) signals, deterministic, statistical, and hybrid approaches channel models may be used depending to the specific application scenarios \cite{han2022terahertz}. Considering the VLP, for example, its RSS in mW can be expressed as:
\begin{equation}\label{Pr_RSSR}
{{P}_{r}}=\frac{U}{{{d}^{m+3}}},
\end{equation}
where $d$ is the distance between the transmitter and the receiver, and $U$ is a parameter related to the transmit power, the configurations of the transmitter and the receiver, and the emergence and incidence angles of the transmission link. Therefore, when the system parameters are known, $U$ can then be calculated, and $d$ can be determined from measured ${{P}_{r}}$ according to (\ref{Pr_RSSR}). 

Based on RSS, two types of positioning algorithms can be used: (i) the fingerprinting algorithm, where the collected data from various known locations is stored in a database. To localize a device, it measures in real-time the RSS values from the surrounding APs in real-time, and compares them with fingerprints stored in the device. Common matching techniques include the following methods \cite{zafari2019survey}: probabilistic, artificial neural networks, k-nearest neighbor, and support vector machine. (ii) The second type is proximity, which is a simple matching strategy that estimates the location of the device to be the same as that of the nearest access point (AP). However, its accuracy is limited and heavily dependent on the density of APs. RSS-based positioning is particularly advantageous due to its low hardware requirements and ease of implementation, making it a preferred method for situations where advanced PSs may not be practical or cost-effective. It is widely used in a variety of environments, from commercial settings for tracking customers to industrial settings for monitoring assets. However, RSS also faces some challenges. In complex indoor environments, factors such as multipath propagation can significantly distort the RSS values, resulting in inaccuracies in positioning. Additionally, environmental dynamics such as human movement and changes in interior layout negatively affect RSS-based methods.


\emph{CSI}: 
CSI refers to the fine-grained characteristics of the wireless propagation path such as attenuation and phase shift, which is crucial in both data transmission and positioning. In contrast to RSS which measures the average amplitude of the signal across its entire bandwidth and aggregates signal strength from all antennas, CSI measures both the amplitude and the phase of each carrier frequency \cite{yang2013rssi}. Zheng \emph{et al.} \cite{zheng2023novel} proposed a support vector machine model for an NLOS-based system based on CSI amplitude, which outperformed the Rician-K and Skewness NLOS detection methods. Moreover, both the channel impulse response (CIR) and the channel frequency response (CFR), which are two variations of CSI techniques, are commonly used in multipath environments for different PSs including geometrical methods \cite{mendrzik2018harnessing}, fingerprinting \cite{sobehy2020csi,fan2021single}, or ML-based method \cite{ferrand2020dnn,de2020csi,B}.

It is generally acknowledged that CSI provides a high level of granularity for precise location estimation and is more robust in multipath and NLOS scenarios than RSS-based PSs. Additionally, CSI provides a wealth of information that can be used by ML algorithms to further enhance PA. However, in dense, cluttered indoor environments, signal reflection, and occlusion can have a significant impact on CSI, which further affects the PA. Moreover, the calibration of the CSI-based PSs can also be laborious in site surveying. Table \ref{tab1} summarizes and compares the key features of all measurements.

\begin{table*}
\setlength{\tabcolsep}{3.5pt}
\centering
\caption{Positioning Measurement Comparison}
\label{tab1}
\fontsize{9}{10}\selectfont
\begin{threeparttable}

\begin{tabular}{c|c|c|c|c|c|c|c}

\hline
Measurement & Accuracy & Cost  &Real-time  &\makecell{Synchro-\\nization}  &\makecell{Privacy and\\ Security} & Computation &Comments\\
\hline
TOA	& \makecell{Centimeter to \\Sub-meter-level }& ● &Hard& Yes& $\bigodot$& $\bigodot$& \makecell{Highly biased from the \\environment,
additional hardware\\ may be required.
}\\ \hline
TDOA& \makecell{Centimeter to \\Sub-meter-level }	& ● & Hard & Yes& $\bigodot$ &	$\bigodot$& \makecell{Performance is degraded \\in the NLOS condition.}\\ \hline
AOA &	\makecell{Sub-meter to \\meter-level} &	● &  Soft	&  No &$\bigodot$ &● & \makecell{LOS signal propagation and \\additional hardware is required \\(e.g., antenna arrays).}\\ \hline
POA &\makecell{Sub-meter to\\ meter-level }&$\bigodot$& Soft& Yes& $\bigodot$ &● & \makecell{LOS signal propagation is \\required .}\\ \hline
RSS	& \makecell{Meter-level or\\ lower }& $\bigcirc$& 	Hard	&No	&$\bigcirc$& $\bigcirc$ & \makecell{PA is susceptible to \\obstacles and multipath \\propagation.}\\ \hline
CSI	& \makecell{Centimeter-level\\ or higher }	& ● & Hard & No &● &$\bigodot$ & \makecell{Special hardware is required.}\\
\hline
\end{tabular}

   \begin{tablenotes}
    \centering
    \footnotesize
		\item {$^{\rm *}$We roughly categorize KPI values in the table into 3 levels, e.g. $\bigcirc$: Low, $\bigodot$: Medium, ●: High.  }
     \end{tablenotes} 
\end{threeparttable} 

\end{table*}

\section{Key Enabler Techniques} 

\subsection{Machine learning} 
ML enables computers to analyze data (i.e., user behavior data, wireless network, and environmental data) for IPS. In particular, ML has two key roles in PSs. (i) ML can extract the positioning features from wireless pilot signals to build a relationship between them and user positions. Here, ML algorithms can be viewed as black boxes with the inputs being wireless pilot signals (in the time or frequency domain) or features (i.e., RSS) of CSI signals and the output being the user’s position. As opposed to traditional methods such as TOA methods \cite{chugunov2020toa}, which manually extract user positioning features (i.e., the distance between the access point and the user), ML-based PSs can automatically extract user positioning features. Hence, ML-based PSs can extract more features from signals in order to determine a user’s position. The ML methods, however, require several labeled data points for training, which is a time-consuming and labor-intensive process. (ii) ML extracts CSI from pilot signals and the extracted CSI, which will be used for traditional positioning methods. For example, one can use ML methods to (i) determine whether the transmission link is LOS or NLOS; (ii) predict the arrival time of the signal; (iii) estimate the distance between the BS and the user; and (iv) estimate the angle differences between two antennas. The ML methods, as opposed to traditional methods \cite{han2020indoor} not being able to extract CSI features accurately in NLOS-based systems, is capable of analyzing the hidden wireless environmental features, resulting in accurate CSI features that can be used to perform traditional positioning. In addition, ML methods for CSI feature extraction can be trained by using the simulated data, which reduces the overhead associated with generating labeled data.

Next, we discuss several recent works on the use of ML algorithms for both user positioning and CSI feature extraction. Current works \cite{raes2020experimental}\cite{cerar2021improving}\cite{cheng2022improved}\cite{bai2019dl} have studied the use of multilayer perceptron, convolutional neural networks (CNNs), recurrent neural networks (RNNs), generative models, and attention-based networks (i.e., transformer) to process CSI data and directly output user positions. Note, (i) a CNN is used when the CSI data can be viewed as an image (i.e., CSI data generated from multiple antennas); and (ii) RNNs are used when the CSI data are time-dependent, while attention-based networks are used to extract CSI features that can significantly contribute to estimation of the user position. Meanwhile, some current works \cite{nabati2023real}\cite{yin2019csi} have studied the use of previously mentioned NNs for CSI feature prediction. In \cite{nabati2023real}, the deep neural network (DNN) method was proposed to learn the distribution of known RSS samples by estimating a user’s location by comparing the similarity of online RSS samples with the reference fingerprints. In \cite{yin2019csi}, three DNN models were developed using pre-proposed data for training. Following training, a subset of samples is selected from the training set for assessing the models, which are then employed during the testing phase to predict real-time CSI data.

To further improve the user position and CSI feature prediction accuracy, current works \cite{wang2018deepml}\cite{li2021toward}\cite{chen2018indoor} also investigated the use of ML to analyze multi-modal data (i.e., camera images, CSI data, earth magnetic field readings) for user positioning. In \cite{wang2018deepml}, a system based on deep long-short-term memory (LSTM) was proposed for indoor positioning using magnetic and light sensors embedded in smartphones. In \cite{li2021toward}, the amplitude information extracted from the CSI together with the calibrated phase information as fingerprints were used to train a DNN-based regression model in order to estimate the target location. The authors \cite{chen2018indoor} utilized a CNN-based image retrieval strategy that represented the scene by CNN features and matches the query image with database images. In \cite{B}, a VLC-IPS with a camera-based receiver was proposed, where the receiver’s position is precisely estimated based on the decoded block coordinate and backpropagation ANN with a mean PA of 1.49 cm.

\subsection{Large models}
Large models belong to the field of ML. To distinguish large models from positioning methods based on traditional ML methods, we introduce them separately due to their immense potential. The traditional ML-based method has two issues: (i) largely relying on labeled data, resulting in the need for a significant amount of manual labor; and (ii) limited adaptability to new environments. Large models are expected to alleviate these problems. First, large models can analyze the wireless network’s environment, temporal and situational with exceptional accuracy, thus accurately identifying users’ locations. Multimodal large language models, through the integration of multimodal data, are expected to parse and comprehend various information in the environment, including RF-based feedback signals, visual gestures, inertial measurement unit (IMU) motion sensor data, and 3D maps\cite{xu2024large}, thus introducing sensing and prediction of the surrounding environment in complex settings. Second, multimodal large models should be able to understand the connections between RF, visual, and inertial data among other modal data types, thereby reducing the need for labeled RF data and the manual labor costs associated with data annotations. Moreover, considering the generative capabilities of multimodal large models, it is anticipated that limited wireless data will be able to generate super-resolution 3D images of the surrounding environment. Incorporating 3D image data with the RF data may provide a better understanding and prediction of user behavior \cite{bariah2024large}, resulting in better proactive positioning, beamforming, power distribution, switching, and spectrum management.

Overall, large models have a wide range of applications in positioning. To benefit from the potential of large models in PSs, it is essential to enhance their ability to understand and predict environments, as well as unlock their ability to align and comprehend multimodal data.

\subsection{Adaptive filter} Adaptive filter refers to a digital filter that dynamically adjusts its coefficients to adapt to changing properties of the signal or the environment in which it operates. It is widely used in fusion positioning due to its ability to iteratively optimize estimations as well as robustness to the noise and interference \cite{fusion-ok}. Different sources of data may have varying levels of accuracy and features, which change over time due to environmental factors like signal obstruction or sensor errors. Inertial navigation methods, for example, provide continuous estimation of target orientations and positions but suffer from the problem of cumulative errors \cite{INS_error}. PSs based on the UWB can provide valuable precious position observations, albeit with the limitation of intermittent output. UWB can assist in the correction of inertial navigation errors, while inertial navigation can provide stable positioning services when UWB fails. Therefore, exploiting the complementary nature of deeply fusing diverse positioning methods based on adaptive filters and fusing them to obtain more accurate estimates of position and orientation has received considerable attention \cite{compl}. 
The two most employed filters in PSs are the Kalman filter (KF) and the particle filter (PF) \cite{kf-pf}.

\subsubsection{Kalman Filter based methods}
 Kalman filter is designed to process a sequence of measurements observed over time, which may contain statistical noise and various inaccuracies. It generates more precise estimates of unknown variables than those derived from a single measurement. This is achieved by estimating a joint probability distribution of the variables for each timeframe, thereby enhancing the accuracy of the output. This technique involves the acquisition of two sets of data: the estimation from the previous time step and the real-time measurement \cite{kalman_true}. As a result of combining these two sets of data in real-time estimation, the obtained estimation represents the transition process of the system state. This approach addresses the challenging task of estimation in non-stationary random processes. 
 However, the initial implementation of KF primarily relied on the state equation, making it only applicable to linear systems. In subsequent research, various improved KF techniques have been developed for the optimization of estimates in nonlinear systems. Among them, the most representative ones are the extended Kalman filter (EKF) and the unscented Kalman filter (UKF).

 The EKF introduces Jacobian matrices to address the challenges in nonlinear systems by means of local linearizing. Feng \emph{et al.} \cite{IMU-UWB} proposed an integrated IPS using EKF, demonstrating that the proposed algorithm can significantly reduce the complexity and costs of base station deployment. In \cite{BLE-AFEKF}, an adaptive feedback extended Kalman filter (AFEKF) algorithm was proposed to fuse Bluetooth low energy (BLE) and pedestrian dead reckoning (PDR), in which the range measurement is deeply fed back to the estimated position at the next moment. Experimental results showed that the AFEKF algorithm improves the accuracy by 23.4$\%$ compared with the classical EKF algorithm.

The UKF scheme combines the unscented transform (UT) with KF framework, thereby making the equations of a nonlinear system compatible with the standard KF framework. In \cite{Data-Fusion}, a multisensor fusion technology based on UKF was used to avoid the issue of neglecting the high-order terms of the nonlinear observation equations of UWB and IMU, which have the potential to improve PA. In \cite{Adaptive-Maximum}, an adaptive maximum correntropy unscented Kalman filter (AMCUKF) was proposed to fuse IMU and UWB data. Using the maximum correntropy criterion, the algorithm suppresses the non-Gaussian noise, thus improving the PA and robustness in complex environments.

In addition to the enhancements to the KF mentioned above, other fusion solutions based on KF have been introduced. The authors in \cite{Federated} proposed an adaptive federated Kalman filter (AFKF) algorithm, where the sharing factors of information fusion and distribution in the FKFr are adaptively adjusted based on the information of sub-filters. The results showed that the PA is improved by more than 10$\%$  compared with other FKF algorithms. In \cite{Ensemble-Transform-KF}, an enhanced ensemble transform Kalman filter (ETKF) was proposed, which fused the predicted position by a PDR and the positional measurement by RSS fingerprinting, thereby estimating the user position based on the ensemble transformation. The experimental results showed that the enhanced ensemble ETKF can achieve higher PA than ETKF and other ensemble-based KFs\cite{Ensemble-Transform-KF}.

\subsubsection{Particle Filter based methods}
PF is a nonparametric Bayesian filter algorithm based on Monte Carlo methods and is employed for the estimation of states in hybrid PSs. Compared with the KF algorithm, a unique feature of the PF algorithm is its sampling approach, which utilizes a set of randomly sampled particles with the associated weights to approximate the posterior distribution of the state \cite{pf_true}. Note that by (i) relaxing the constraints of linearity and Gaussianity, it is possible to handle nonlinear models and non-Gaussian noise distributions; and (ii)  adjusting the weights and positions of particles, the algorithm yields an estimate of the state of the system, i.e., the position of the user. The selection and computation of the weights depend on the PSs to be fused. In \cite{data-driven}, a feasible method utilizing particle filter to fuse data-driven inertial navigation and BLE was proposed for indoor positioning. The proposed fusion algorithm reduced the mean positional error by more than 25$\%$  compared with Bluetooth-based positioning.

\begin{table*}[htbp]
 \centering
\caption{Adaptive Filter Based Fusion Positioning Comparison}
\label{tab_adaptive}
\renewcommand{\arraystretch}{1.5}
\setlength{\tabcolsep}{3.5pt}
\scriptsize
\begin{tabular}{|c|c|c|c|c|c|c|c|c|c|}
\hline

\multirow{2}{*}{Base Filter} & \multirow{2}{*}{System}                       & \multirow{2}{*}{Positioning Algorithm}                                                                  & \multirow{2}{*}{Filter Algorithm} & \multicolumn{6}{c|}{Evaluation Framework}                                                                                                                                                                   \\ \cline{5-10} 
                             &                                               &                                                                                                          &                                   & \multicolumn{1}{c|}{Availability} & \multicolumn{1}{c|}{Accuracy}                 & \multicolumn{1}{c|}{Cost} & \makecell{Environment\\-friendly} & \multicolumn{1}{c|}{Portability} & Instantaneity \\ \hline
Kalman Filter                 & \cite{IMU-UWB}               & Extended Kalman Filter (EKF)                                                                             & UWB + INS                         & \multicolumn{1}{c|}{√}           & \multicolumn{1}{c|}{Centimeter-level (cm)}    & \multicolumn{1}{c|}{High} & \multicolumn{1}{c|}{√}                    & \multicolumn{1}{c|}{Low}         & √             \\ \hline
Kalman Filter                 & \cite{Data-Fusion}           & Unscented Kalman Filter (UKF)                                                                            & UWB + INS                         & \multicolumn{1}{c|}{√}           & \multicolumn{1}{c|}{Centimeter-level (cm)}    & \multicolumn{1}{c|}{High} & \multicolumn{1}{c|}{√}                    & \multicolumn{1}{c|}{Low}         & √             \\ \hline
Kalman Filter                 & \cite{Adaptive-Maximum}      & \makecell{Adaptive Maximum Correntropy \\ Unscented Kalman filter (AMCUKF)} & UWB + INS                         & \multicolumn{1}{c|}{√}           & \multicolumn{1}{c|}{Sub-meter to meter-level} & \multicolumn{1}{c|}{High} & \multicolumn{1}{c|}{√}                    & \multicolumn{1}{c|}{Low}         & √             \\ \hline
Kalman Filter                 & \cite{Federated}             & \makecell{Adaptive Federated\\Kalman filter (AFKF)}                                                                 & UWB + INS                         & \multicolumn{1}{c|}{√}           & \multicolumn{1}{c|}{Centimeter-level (cm)}    & \multicolumn{1}{c|}{High} & \multicolumn{1}{c|}{√}                    & \multicolumn{1}{c|}{Low}         & √             \\ \hline
Kalman Filter                & \cite{li2019indoor}             & Extended Kalman Filter (EKF)                                                                             & UWB + INS                         & \multicolumn{1}{c|}{√}           & \multicolumn{1}{c|}{Sub-meter to meter-level} & \multicolumn{1}{c|}{High} & \multicolumn{1}{c|}{√}                    & \multicolumn{1}{c|}{High}        & √             \\ \hline
Kalman Filter                 & \cite{Ensemble-Transform-KF} & \makecell{Ensemble Transform\\Kalman filter (ETKF)}                                                                  & WIFI + INS                        & \multicolumn{1}{c|}{√}           & \multicolumn{1}{c|}{Meter-level}              & \multicolumn{1}{c|}{Low}  & \multicolumn{1}{c|}{√}                    & \multicolumn{1}{c|}{Low}         & ×             \\ \hline
Particle Filter              & \cite{Tight-couple}          & Particle Filter (PF)                                                                                     & WIFI + INS                        & \multicolumn{1}{c|}{√}           & \multicolumn{1}{c|}{Sub-meter to meter-level} & \multicolumn{1}{c|}{High} & \multicolumn{1}{c|}{×}                    & \multicolumn{1}{c|}{Low}         & ×             \\ \hline
Particle Filter              & \cite{Federate}              & Federated Particle Filter (FPF)                                                                          & WIFI + PDR                        & \multicolumn{1}{c|}{√}           & \multicolumn{1}{c|}{Sub-meter to meter-level} & \multicolumn{1}{c|}{Low}  & \multicolumn{1}{c|}{√}                    & \multicolumn{1}{c|}{High}        & ×             \\ \hline
Particle Filter              & \cite{max-like}              & \makecell{Maximum Likelihood\\Particle Filter (MLPF)}                                                                & WIFI + INS                        & \multicolumn{1}{c|}{√}           & \multicolumn{1}{c|}{Sub-meter level}          & \multicolumn{1}{c|}{High} & \multicolumn{1}{c|}{√}                    & \multicolumn{1}{c|}{Low}         & √             \\ \hline
Particle Filter              & \cite{2023spotter}              & Particle Filter (PF)                                                                                     & WIFI + BLE                        & \multicolumn{1}{c|}{√}           & \multicolumn{1}{c|}{Meter-level}              & \multicolumn{1}{c|}{High} & \multicolumn{1}{c|}{√}                    & \multicolumn{1}{c|}{Low}         & ×             \\ \hline
Kalman Filter                & \cite{9535887}                  & Unscented Kalman Filter (UKF)                                                                            & \makecell{WIFI + BLE\\+ PDR }                     & \multicolumn{1}{c|}{√}           & \multicolumn{1}{c|}{Meter-level}              & \multicolumn{1}{c|}{High} & \multicolumn{1}{c|}{√}                    & \multicolumn{1}{c|}{Low}         & √             \\ \hline
Particle Filter              & \cite{tc-wifi}                  & Extended Kalman Filter (EKF)                                                                             & WIFI + PDR                        & \multicolumn{1}{c|}{√}           & \multicolumn{1}{c|}{Meter-level}              & \multicolumn{1}{c|}{High} & \multicolumn{1}{c|}{√}                    & \multicolumn{1}{c|}{Low}         & ×             \\ \hline

Kalman Filer                 & \cite{BLE-AFEKF}             & \makecell{Adaptive Feedback Extended\\Kalman filter (AFEKF)}                                                        & BLE + PDR                         & \multicolumn{1}{c|}{√}           & \multicolumn{1}{c|}{Sub-meter to meter-level} & \multicolumn{1}{c|}{Low}  & \multicolumn{1}{c|}{√}                    & \multicolumn{1}{c|}{High}        & √             \\ \hline

Particle Filter              & \cite{data-driven}           & Particle Filter (PF)                                                                                     & BLE + INS                         & \multicolumn{1}{c|}{√}           & \multicolumn{1}{c|}{Meter-level}              & \multicolumn{1}{c|}{Low}  & \multicolumn{1}{c|}{√}                    & \multicolumn{1}{c|}{High}        & ×             \\ \hline

Kalman Filter                & \cite{huang2019hybrid}          & Extended Kalman Filter (EKF)                                                                             & BLE + PDR                         & \multicolumn{1}{c|}{√}           & \multicolumn{1}{c|}{Sub-meter to meter-level} & \multicolumn{1}{c|}{High} & \multicolumn{1}{c|}{√}                    & \multicolumn{1}{c|}{Low}         & ×             \\ \hline
Kalman Filter                & \cite{chen2021pre}              & Extended Kalman Filter (EKF)                                                                             & \makecell{Acoustic Ranging\\+ PDR}            & \multicolumn{1}{c|}{√}           & \multicolumn{1}{c|}{Sub-meter to meter-level} & \multicolumn{1}{c|}{Low}  & \multicolumn{1}{c|}{√}                    & \multicolumn{1}{c|}{Low}         & ×             \\ \hline

Particle Filter              & \cite{he2023vehicle}            & Particle Filter (PF)                                                                                     & VLP + INS                         & \multicolumn{1}{c|}{√}           & \multicolumn{1}{c|}{Centimeter-level (cm)}    & \multicolumn{1}{c|}{Low}  & \multicolumn{1}{c|}{√}                    & \multicolumn{1}{c|}{High}        & √             \\ \hline
Kalman Filter                & \cite{2019liang-vlc}            & Extended Kalman Filter (EKF)                                                                             & VLP + INS                         & \multicolumn{1}{c|}{√}           & \multicolumn{1}{c|}{Sub-meter level}          & \multicolumn{1}{c|}{High} & \multicolumn{1}{c|}{√}                    & \multicolumn{1}{c|}{High}        & √             \\ \hline
Kalman Filter                & \cite{carreno2020opportunistic} & Extended Kalman Filter (EKF)                                                                             & VLP + PDR                         & \multicolumn{1}{c|}{√}           & \multicolumn{1}{c|}{Sub-meter to meter-level} & \multicolumn{1}{c|}{Low}  & \multicolumn{1}{c|}{√}                    & \multicolumn{1}{c|}{High}        & √             \\ \hline
Particle Filter              & \cite{li2017fusion-pf}          & Particle Filter (PF)                                                                                     & VLP + PDR                         & \multicolumn{1}{c|}{√}           & \multicolumn{1}{c|}{Sub-meter level}          & \multicolumn{1}{c|}{Low}  & \multicolumn{1}{c|}{√}                    & \multicolumn{1}{c|}{High}        & ×             \\ \hline
Kalman Filter                & \cite{li2017fusion-kf}          & Extended Kalman Filter (EKF)                                                                             & VLP + PDR                         & \multicolumn{1}{c|}{√}           & \multicolumn{1}{c|}{Sub-meter level}          & \multicolumn{1}{c|}{Low}  & \multicolumn{1}{c|}{√}                    & \multicolumn{1}{c|}{High}        & ×             \\ \hline
Kalman Filter                & \cite{guan2021robust}           & Extended Kalman Filter (EKF)                                                                             & VLP + INS                         & \multicolumn{1}{c|}{√}           & \multicolumn{1}{c|}{Centimeter-level (cm)}    & \multicolumn{1}{c|}{High} & \multicolumn{1}{c|}{√}                    & \multicolumn{1}{c|}{Low}         & √             \\ \hline

\end{tabular}

\end{table*}

The current research primarily focuses on enhancing the weight strategy and modifying the structure of filters. The authors in \cite{firefly} proposed an optimized particle filter algorithm that fused PDR and geomagnetic positioning by introducing a firefly algorithm to optimize PF, thereby enhancing particle updating and target state detection. Compared with the conventional particle filter, the PA was improved by 120$\%$. In \cite{Federate}, a federated particle filter (FPF) with information sharing was proposed to fuse PDR and WiFi. The system is comprised of multiple sub-filters and a primary filter. The observed data input was initially optimized for the corresponding sub-filters. Subsequently, the obtained output was applied to the primary filter for the final estimation. The experimental results demonstrated that the proposed method can effectively control the accuracy to within approximately 1 m. The authors in \cite{Tight-couple} proposed TrackInFactory, a solution based on PF that fuses INS and WiFi information in a novel way. The scheme dynamically updates the particles’ weights using a new and reliable metric that defines the confidence of each position estimate, with a mean error of 0.81 m. Also, in \cite{max-like}, a novel maximum likelihood particle filter was proposed to ensure that all particles are efficiently used. The performance of the algorithm exceeded the requirements of the 5G NR Release 16 standard from 3GPP. In \cite{VISEL}, the authors developed a high-precision PS that completed an enhanced particle filter with an adaptive reassignment of weights to different positioning modules. The system outperformed the current state-of-the-art PSs and achieved an average PA of 0.4 m.

In summary, adaptive filters have gained widespread application in fusion positioning due to their ability to autonomously update filter coefficients depending on the environment in which they are used. There are, however, some challenges associated with adaptive filters including the coefficient adjustment delays and slow convergence rates, which makes them less suitable for real-time data fusion tasks with stringent timing requirements\cite{delay}. Table \ref{tab_adaptive} summarizes the current research and provides an evaluation of the attributes of the fusion PSs based on adaptive filter \cite{ IMU-UWB, BLE-AFEKF, Data-Fusion, Adaptive-Maximum, Federated, Ensemble-Transform-KF, data-driven, Tight-couple, Federate, max-like, 2023spotter, 9535887, tc-wifi, huang2019hybrid, chen2021pre, li2019indoor, he2023vehicle, 2019liang-vlc, carreno2020opportunistic, li2017fusion-pf, li2017fusion-kf, guan2021robust}.

\subsection{Reconfigurable Intelligent Surface} 
RIS is a plane composed of numerous tiny antenna components, which can be programmatically controlled to dynamically modify the propagation characteristics of electromagnetic waves (i.e., amplitude, phase, and polarization) \cite{wu2019intelligent,wu2019towards,long2021promising}. RIS optimizes the performance of wireless communication networks by efficiently controlling wireless signals through altering the electromagnetic wave propagation environment \cite{chen2022reconfigurable}. RIS operates on the principle of electromagnetic wave reflections. Specifically, when electromagnetic waves, such as wireless signals, encounter the RIS, each scattering element of RIS independently adjusts the phase and amplitude of the reflected waves. By precisely adjusting these parameters, RIS can change the propagation direction of electromagnetic waves, and concentrate or disperse energy, thus controlling the propagation of signals in specific directions.

In PSs, RIS generally plays two roles: (i) as passive reflectors, which are most used \cite{wymeersch2019radio}; and (ii) active reflectors (i.e., active transceivers)  \cite{chen2022reconfigurable}. When RIS acts as reflectors, it can create additional signal propagation paths to bypass blocking and shadowing,  thus introducing extra degrees of freedom in the design of PSs \cite{jian2022reconfigurable}. In \cite{he2020large}, a reflector-based RIS was introduced in a mmWave multiple-input multiple-output (MIMO)-based PS. Also introduced were Fisher information matrix (FIM) and Cramer–Rao lower bound (CRLB) for the standard deviation of the positioning estimation error as well as the orientation estimation error, which demonstrated that the proposed PS is superior to the traditional PS. The authors in \cite{liu2021reconfigurable} used reflector-based RIS in the scenarios with no LOS paths, and derived the FIM and CRLB in order to estimate the absolute position of the mobile station. By optimizing the reflect beamforming design to minimize CRLB, the PA was improved by the decimeter level or even the centimeter level\cite{zhang2021metalocalization,nguyen2020reconfigurable}. By acting as transmitters \cite{basar2019wireless} or receivers \cite{taha2021enabling}, the RIS can be operated as a reconfigurable lens in PSs.

The advantages of applying RIS in PSs are as follows: (i) significantly enhanced PA by adjusting signal propagation paths \cite{hu2017cramer}; (ii) enhancing the coverage area by smartly reflecting signals to avoid obstacles, thereby establishing adaptive virtual LOS connections in areas with poor coverage or blind spots \cite{zeng2020reconfigurable}; and (iii) cost-effective,  using reflective components, miniature antennas, and diodes. The challenges of RIS in PSs, however, are in the design and implementation complexity, highly precise control, standardization, and compatibility \cite{wymeersch2019radio}. In addition, the RIS technology lacks unified standards \cite{liu2022path}, and more research works need carrying out on protocols\cite{croisfelt2022random}.

\subsection{Software Defined Network (SDN)} 
SDN represents a new paradigm of network architecture designed to enhance network flexibility, manageability, and programmability \cite{kobo2018fragmentation,yassein2017combined}. The fundamental idea of SDN is to separate the network control layer from the data forwarding layer \cite{kobo2017survey}, which allows more agile handling of network traffic and policies \cite{cloete2019review}. In conventional networks, each network device, such as switches and routers, possesses its own control logic and forwarding functions. As a result of SDN, network management is simplified and optimized by abstracting the control logic  (which determines how and where data is forwarded) from physical devices and centralizing it into a single point of control,i.e., the SDN controller \cite{cloete2019review,junfeng2010mds,kobo2017towards,pritchard2017security,kim2013improving}.

In wireless sensor networks, SDN can enhance the efficiency and accuracy of positioning services \cite{zhu2016software}. In PSs, SDN can be used with either gainful methods or ungainful methods \cite{shahal2023review}, where in the former the focus is on enhancing PA and reducing energy consumption. Kim \emph{et al.} \cite{zhu2016software} proposed an SDN-based positioning node selection algorithm that used a linear least square algorithm and RSS measurements to implement Euclidean position estimation. Simulation results showed up to 45\% increase in PA. In \cite{zhu2017sdn}, a centralized anchor scheduling scheme was proposed, which used the SDN controller to broadcast messages among nodes and localized mobile agents. Based on simulation results with a 14,400 $\rm{m^{2}}$ sensor field with 200 randomly placed anchor nodes and 10 mobile agents, it was shown that the scheme reduced the number of active anchor nodes and reduced the PA with a significant reduction in the energy consumption, thereby increasing the network lifetime. Some similar works can be found in\cite{zhu2018node,zhu2017localisation,zhu2021improving}. In ungainful methods, SDN does not typically incorporate the computational requirements of positioning in the control-plane \cite{cloete2019comparison}. Instead, they explore the potential of combining SDN with positioning.

SDN can enhance various aspects of positioning, such as reducing energy consumption and improving accuracy \cite{de2014smart}. Specifically, SDN can not only provide an energy-efficient method for managing sensors but also manage networks, thereby reducing convergence time. Due to these two characteristics, SDN can reduce energy consumption and reduce positioning latency in PSs \cite{de2014smart}. Based on high centrality and global perspective on positioning nodes \cite{cloete2019review}, SDN has improved PA \cite{shahal2023review,zhu2016software,zhu2017localisation} is attributed to its.

\section{Technologies and Solutions}

\subsection{Celluar Networks}
Positioning has always been an integral component of standardized 3GPP technologies. In 3GPP 5G New Radio (NR), UEs are provided with enhanced positioning capabilities. In terms of frequency bands, NR operates over a wide frequency spectrum of below 6 GHz and the lower mmWave range of 24.25 GHz. This allows NR to leverage a wide signal bandwidth to achieve higher PA with timing measurements. A 5G enabler with higher data throughput and coverage areas, massive antenna arrays, and beamforming techniques can also be leveraged to locate UEs through accurate angular measurements.

3GPP 5G NR has supported positioning features since its inception in Release-15. However, Release 15 positioning support is limited to the so-called RAT (Radio Access Technology)-independent positioning methods (i.e., using signals from the UE’s various sensors and WiFi/Bluetooth receivers) and LTE-based positioning. 3GPP 5G NR Release-16 introduces native 5G positioning signals and extends the standardized positioning capability beyond those defined in 4G LTE. Release-16 specifies a range of PSs to satisfy the needs of regulations, such as FCC’s e911 emergency calls requirements, and commercial use cases, such as emergency calls, indoor factories, and vehicle-to-everything (V2X). The target requirement for commercial use cases is to achieve a 2D positioning accuracy of less than 3 m and 10 m for 80\% of UEs in indoor and outdoor scenarios, respectively. The regulatory requirement mandates a 2D PA of 50 m for both indoor and outdoor applications. The PSs include those using the timing measurements between the UE and multiple transmission-reception points: downlink or uplink TDoA and multi-cell round trip time (multi-RTT). In terms of reference signals, the uplink-sounding reference signal (UL- SRS) for positioning and the downlink positioning reference signal (DL-PRS) were introduced in Release-16. Both can be configured with a bandwidth in the range of 24 to 276 PRBs in steps of 4 PRBs. This provides a large bandwidth of up to 100 MHz for a 30 kHz subcarrier spacing in FR1, and up to 400 MHz for a 120 kHz subcarrier spacing in FR2. As a result of large bandwidth, timing measurement can be much more precise than that of LTE. Additionally, positioning methods are defined to leverage angular measurements from antenna arrays, namely the downlink angle of departure (DL-AoD) and the uplink angle of arrival (UL-AoA).

3GPP Release-17 addresses the stringent requirements of new applications and industry verticals, including increased accuracy and lower latency, while maintaining high integrity and reliability\cite{3g2021study}. For general commercial use cases, the target requirements for 90\% of UEs are horizontal and vertical PAs of \textless 1 and \textless 3  m, respectively. For industry IoT (IIoT) use cases (e.g., factory automation), the target requirements for 90\% of UEs are horizontal and vertical PAs of  \textless 0.2 and \textless 1 m, respectively. Release-17 specified numerous enhancement features to satisfy the tight requirements\cite{sfsou2023}. These include methods to mitigate transmission and reception timing errors at the UE and gNB; methods to improve angular measurements for DL-AoD and UL-AoA; LOS -NLOS indicator; positioning of UEs in the inactive state; on-demand transmission and reception of DL PRS; and GNSS positioning integrity determination.

In 2023, the work on 3GPP Release-18 for positioning is being carried out\cite{Soeai2022}, where 5G NR positioning features are further enhanced, including:
\begin{itemize}
\item[$\bullet$] Two methods are specified for achieving higher PA: (i) increasing the transmission/reception bandwidth of the DL and UL reference signals for positioning by bandwidth aggregation of intra-band contiguous carriers; and (ii) using the NR carrier phase measurements to achieve centimeter-level PA, similar to GNSS carrier case positioning defined for outdoor applications.
\end{itemize}

\begin{itemize}
\item[$\bullet$] Sidelink (UE-to-UE) positioning is supported in all coverage scenarios (in-coverage, partial coverage and out-of-coverage) with a focus on V2X and public safety use cases.
\end{itemize}

\begin{itemize}
\item[$\bullet$] Low power high accuracy positioning (LPHAP) is supported for IIoT use cases such as massive asset tracking and automated guided vehicles (AGV) tracking in factories. The emphasis is on lower UE power consumption while achieving a target accuracy of \textless 1 m, where the device battery life is expected to last from 6 months to a year.
\end{itemize}

\begin{itemize}
\item[$\bullet$] PA enhancement features are introduced in Redcap UEs, to deliver high-accuracy positioning even for devices with a limited RF bandwidth.
\end{itemize}

\begin{itemize}
\item[$\bullet$] Positioning integrity is supported for mission-critical use cases that rely on positioning estimates and uncertainty estimates. The integrity of RAT-dependent positioning methods provides a measure of trust in the accuracy of the position-related data as well as the capability to provide timely alerts when the accuracy may deteriorate beyond acceptable levels.
\end{itemize}

\begin{figure}
    \centering
    \includegraphics{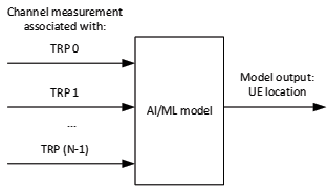}
    \caption{Direct AI/ML positioning.}
    \captionsetup{justification = centerlast}
    \label{fig:my_label1}
\end{figure}

\begin{figure}
    \centering
    \includegraphics{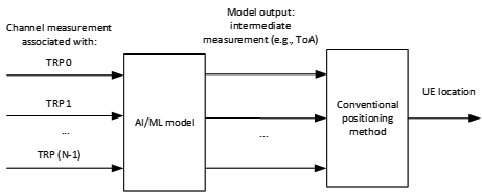}
    \caption{AI/ML assisted positioning.}
    \captionsetup{justification = centerlast}
    \label{fig:my_label2}
\end{figure}

In parallel to the Release-18 work item for positioning, a study on ML-based positioning is being conducted from May 2022 to November 2023, which is a representative use case of Release-18 study on ML for the physical layer\cite{Soaim2023}. Two approaches are investigated: (i) direct ML positioning shown in Fig. \ref{fig:my_label1}, where the model output is the UE position; and (ii) ML assisted positioning shown in Fig. \ref{fig:my_label2}, where the model output is one or more intermediate measurements (e.g., LOS/NLOS indicator and time-of-arrival) that can be utilized by conventional positioning methods to determine the UE position.

ML-based positioning is designed to target challenging scenarios. For example, in a cluttered factory indoor scenario, where links are scarce and the conventional methods that rely on timing (e.g., multi-RTT) and/or angular measurements (e.g., UL-AoA) tend to fail. For example, the probability of LOS paths is 0.8\% in a factory environment with dense clutter and a high base station height, and the clutter parameter settings are clutter density=60\%, clutter height=6 m, clutter width=2 m.  If using conventional positioning methods, the horizontal PA is around 15.8 m at 90\% CDF, indicating very poor PA. In contrast, when using the ML-based positioning method, a PA as low as 20 cm is achievable, depending on design factors such as ML model input, model architecture and size, and training dataset size. Such excellent PA demonstrates that ML-based positioning is a very worthwhile objective for standards to pursue.

In comparison to conventional (i.e., non-ML) positioning methods, ML-based positioning requires a paradigm shift in the design. With conventional methods, a snapshot of the wireless signals is measured and processed to extract the timing and/or angular information, after which the location of the UE can be estimated by triangulation. With the ML-based positioning methods, an ML model is trained to learn from a large training dataset, where the training dataset contains features that are representative of the target deployment scenario. The quality and quantity of the training dataset significantly affect the PA of the model. To support the ML model, a set of ML life cycle management issues need addressing, including training data collection, model training, model monitoring, and model update.

\subsection{WiFi}

With WiFi, electronic devices are able to connect to a wireless local area network (LAN) via the ISM radio band. In addition to providing high data-rate communication services, WiFi sensing has emerged as an innovative approach in environmental sensing. Positioning is one of the most common tasks for WiFi sensing \cite{hernandez2022wifi}, due to the increasing demands for locating humans and devices in smart environments. There are typically two techniques for WiFi positioning. The first type is RSS-based in order to estimate the devices’ position using WiFi networks. An early RADAR system based on RSS was demonstrated in \cite{bahl2000radar}, followed by a series of other schemes such as Redpin \cite{bolliger2008redpin}, LoCo \cite{biehl2014loco} and OpenWRT \cite{estrada2023wifi}. The second type is CSI, was introduced in Section III, in which finer granularity channel information is provided compared to RSS. CSI measurements are also available for off-the-shelf WiFi cards to develop a simple but accurate “effective SNR” model to predict successful packet delivery for a given transmit configuration\cite{halperin2011tool}. In addition, there are other types of PSs, such as time-based WiFi, which is rather complex due to the measurement of the time delay and sensitivity to channel conditions, thus needing further investigations \cite{becker20235g}.

The latest WiFi standard is 802.11ax, which further enhances wall penetration performance compared to WiFi standards. 802.11ax can provide as large as 80 MHz of bandwidth coarsely corresponding to a resolution of 1.88 m \cite{storrer2021indoor}. 802.11 is based on a new structure of high-efficiency frames, which reduces the subcarrier spacing and includes more subcarriers within the same bandwidth, which is beneficial in positioning. It is expected that IEEE 802.11be (i.e. WiFi 7) will extend the bandwidth to 160 MHz, thus further increasing the range resolution \cite{deng2020ieee}. As a result of greater resolution in the frequency domain, a receiver can distinguish a greater number of multipath components. Through this enhanced discrimination, it is possible to improve the estimation of channel parameters such as the AOA and TOF, which are essential measurements for positioning. Furthermore, some WiFi amendments, such as IEEE 802.11mc, include the fine-time measurement (FTM) protocol. This also motivates time-based WiFi positioning studies \cite{si2020wi}. The ubiquitous availability of WiFi makes it a promising indoor positioning technology, however, it consumes a relatively high amount of power compared to cellular or Bluetooth \cite{isaia2023review}. Moreover, the existing RSS/CSI-based positioning method is based on an extensive dataset. However, RSS/CSI values may change over time (months or years), and adapting to these variations is also a prominent challenge in WiFi.

\subsection{Bluetooth}

Bluetooth is a popular short-range RF technology. Bluetooth low energy (BLE) is a low-power Bluetooth wireless communication standard developed by the Bluetooth special interest group (SIG), which is widely used in current devices. Both BLE and classic Bluetooth operate in the 2.4 GHz band\cite{bt1} and use the Gaussian frequency shift keying modulation scheme. Typically, BLE uses devices such as beacons or Bluetooth positioning tags as transmitters, and devices such as smartphones or Bluetooth gateways as receivers, RSSI is used for estimating the receiver distance from the transmitter in. In these systems, the receiver estimates its distance from the transmitter based on the RSSI value. To achieve positioning with this scheme, at least three beacons are needed for trilateration \cite{bt2}. Additionally, the fingerprint method has also been extensively studied.  Pu \textit{et al.} \cite{bt3} proposed a fingerprint PS using the \textit{k}-nearest neighbor (kNN) classification method, while Nguyen \textit{et al.} \cite{bt4} used an improved weighted kNN and Gaussian process regression to achieve BLE-based positioning. Echizennya \textit{et al.} \cite{bt5} investigated a method to simultaneously detect the location and motion direction of a pedestrian walking in an indoor environment using a trained deep NN with a PA of 0.439 m and an average direction accuracy of 81.2\% in 9 directions.

In addition to RSS-based positioning, Bluetooth 5.1 further proposed a centimeter-level PS based on AoA/AoD algorithms. The AoA algorithm uses positioning tags such as beacons or Bluetooth bracelets as transmitters and positioning base stations as antenna arrays as receivers. In order to achieve positioning,  the positioning tag transmits a signal to the antenna array to generate a phase difference to determine the AOA. An antenna array transmitter is used for the AoD positioning, which determines the signal departure angle for positioning. He \textit{et al.} \cite{bt7} proposed an AoA estimation based on multiple antenna arrays, which improved the AoA estimation accuracy with an average error of less than 3.9° compared with multiple signal classification. Zhu \textit{et al.} \cite{bt8} proposed a CNN Bluetooth indoor positioning algorithm based on hybrid RSSI-AoA with improved PA. Note, generally, the PA of AoA/AoD is higher than that RSSI.

In addition to Bluetooth 5.1, SIG has released Bluetooth 5.2 \cite{bt9} and 5.3 \cite{bt10}, which add new features such as LE isochronous channels, enhanced attribute protocols, LE power control, low-rate connections, enhanced encryption control, enhanced periodic broadcasting, \textit{etc.}. In this way, Bluetooth technology has greatly improved the transmission rate, security and stability of PSs. The main challenge faced by the RSSI-based PS is that it is affected by complex and unpredictable indoor environments and noise., where RSSI values fluctuate greatly, resulting in unstable positioning results. Furthermore, factors such as signal reflection interference and antenna array errors may also affect the performance of AoA/AoD positioning.

\subsection{RFID}

In RFID, objects tagged with RF transceivers are automatically identified and tracked, and the information collected is stored in the computer \cite{panigrahi2015analysis,panigrahi2015low}. An RFID system typically consists of RFID tags and RFID readers with microchips for data storage and antennas\cite{panigrahi2015analysis}. Active RFID tags emit RF signals using their own power sources \cite{shirehjini2012rfid}, whereas passive RFID tags are activated on receiving reader signals \cite{park2009indoor, choi2012indoor}. RFID readers transmit signals to tags and receive responses from them. When a tag is within the reader’s signal range, it responds, allowing the reader to capture and relay the data stored for processing \cite{obeidat2021review,tesoriero2010improving,mi2015performance}.

The RFID protocol standards are broadly classified into three categories based on frequency bands: ISO 14443, ISO 15693, and ISO 18000-6C. ISO 14443 is a protocol for close-range reading, with tag read-write transmission range of 0 to 10 cm. ISO 15693 is designed for longer-range reading, with the tag read-write distances of 0 to 100 cm. ISO 18000-6C supports tag read-write over a range of 0 to 1000 cm, making it suitable for mid-to-long-range applications. ISO 18000- 6Cs defines physical and logical requirements for a passive-backscatter, interrogator-talks-first RFID system operating at 860–960  MHz\cite{zhang2020standards}.

Both active tags and passive tags can be used for positioning. Active tags are mainly used for long-range positioning and object tracking \cite{tesoriero2008using,aggarwal2012rfid,deak2012survey}. In practice, passive RFID tags are more commonly employed in PSs compared to active RFID tags \cite{kim2020review}. Panigrahi \emph{et al.} \cite{panigrahi2015low} proposed a graph-based simulated model for planning the shortest path, where RFID tags were arranged in an equidistance manner in grid-based surroundings to determine the robot’s position. Hybrid RFID systems based on KF, vision, and Bayesian models were also investigated in \cite{yang2018kalman,martinelli2015robot,zhang2018bfvp,hwang2017neural}. The RFID-based system is well suited in indoor environments due to its precise path estimation and low positioning error \cite{panigrahi2022localization}. Active RFID tags are characterized by a greater detection range, and higher power consumption and costs \cite{kim2020review,tesoriero2010improving}. Passive RFID tags are used for short-range, static point positioning in small spaces \cite{chu2011high}. The low costs of passive tags make RFID technologies highly popular in many applications \cite{baha2017accurate}. However, privacy is a concern, especially in passive RFID tags with insufficient computing capability to support cryptographic data protection \cite{basiri2017indoor}.


\subsection{UWB}
UWB is a short-range wireless technology that uses frequencies between 3.1 and 10.6 GHz, which has much wider bandwidth than narrow-band transmissions such as WiFi. The wider bandwidth of UWB allows for better time and distance estimates, resulting in enhanced positioning performance. UWB has been primarily used for positioning purposes in recent years.


PSs using UWB can determine the user’s position utilizing various methods, including both ranging and non-ranging techniques. UWB typically achieves positioning using time-based measurements (i.e., TOA/TDOA), but with strict time synchronization. To eliminate the need of synchronization, a two-way-ranging (TWR) method has been proposed, in which round-trip time (RTT) at the anchor was used to calculate ToF without tag-anchor or anchor-anchor synchronization. With AoA, the location of a tag can be estimated using a single anchor equipped with at least two antennas. Several systems have already been implemented \cite{mazhar2017precise}, EUROPCOM\cite{harmer2008europcom}, and Ubisense\cite{steggles2005ubisense}, while others are being used as experimental testbeds as in Decawave\cite{Link01} and Bespoon\cite{Link02}. In recent years, some scholars have proposed hybrid PSs by fusing UWB with other technologies. For example, in \cite{xue2018research}, a PS based on UWB and dead reckoning algorithm was proposed to overcome the problem of large errors and instability.

Standards and protocols for UWB PSs have been defined by several organizations including IEEE, FiRa, Car Connectivity Consortium (CCC), \textit{etc}. IEEE 802.15.4 is a prominent example, where IEEE 802.15.4a was first released in 2007, and since then it has been revised and improved. In 2020, IEEE 802.15.4z was released with increased integrity and improved accuracy of ranging measurements. Enhancements include additional coding and preamble options, resulting in proportionally fewer zero-valued elements and improved detection. The application of UWB has expanded rapidly, so task groups and organizations (IEEE Task Group 15.4ab, Omlox) have been formed to propose new protocols and standards. It is expected that the cross-system, cross-platform information exchange model between UWB solutions of different vendors and various positioning technologies could be standardized to permit multiple systems to communicate and interoperate with each other, thereby improving context information and resolving positioning errors \cite{coppens2022overview}. Furthermore, standardization of antenna design and new performance metrics are also desirable since improper antenna design may lead to severe pulse distortion and undesired phase center variations. This also motivates AOA UWB positioning studies.

Despite the above advancements, UWB still faces several challenges in practice. For instance, due to high propagation loss and poor penetrating ability, UWB systems are range-limited and require LOS paths between receivers and transmitters, which raises the cost for a greater number of transmitters in indoor environments\cite{sesyuk2022survey}. 

\subsection{mmWave}
mmWave is an emerging wireless technology working in the 30-300 GHz frequency band. Besides higher-rate communications, its short wavelength allows for accurate location estimates and lower location error bounds. Moreover, mmWave propagation characteristics yield higher spatial scanning resolution \cite{mw1}. mmWave positioning algorithms typically make use of signal parameters related to received signal power (RSSI/SNR), time information (ToA/TDoA), angle information (AoA/AoD), CSI, or hybrid approaches to obtain location estimates with high PA. Among these schemes, AoA is the most accurate, due to the exploitation of directional beamforming and antenna arrays in mmWave systems \cite{mw2}.  Li \textit{et al.} \cite{mw3} proposed a novel hybrid dual-polarized antenna array and studied an adaptive AoA and polarization state estimation, showing a significant improvement in SNR. Using both the angle and time has led to improved PA in mmWave systems \cite{mw4}. For example, Jia \textit{et al.} \cite{mw5}  proposed an improved least mean square algorithm to refine AoA estimation, and used a modified multi-path AoA-ToA UKF algorithm to track UE’s position with 2 times angle estimation gain and a centimeter PA using a single AP in an office environment.

In addition, mmWave-based device-free positioning and sensing has also been recognized as an energy-efficient and feasible technology for environmental sensing \cite{mw1}. It typically depends on radar systems that operate over short distances. There are various types of mmWave radars, including pulsed wave radars, frequency shift keying radars, frequency-modulated continuous wave (FMCW) radars, \textit{etc}, \cite{mw6}. FMCW radar is widely used in remote sensing, due to its high resolution, in applications such as human activity detection, object detection, health monitoring, etc. In addition to traditional key processing techniques like micro-Doppler,  KF and ML are being successfully used in mmWave-based radar sensing systems \cite{mw1}. Jin \textit{et al.} \cite{mw7} used a 4-D mmWave radar and a hybrid variational RNN autoEncoder for fall detection of people with a 98\% detection rate. Based on sparse mmWave radar point clouds with a novel DL classifier, Pegoraro \textit{et al.} \cite{mw8} proposed a real-time multi-target tracking and identification system with an identifying accuracy of 91.62\% for up to three mobile subjects in an indoor environment.


In 2012, IEEE 802.11ad standard was released with 60.0 GHz wireless communication features \cite{nitsche2014ieee},  which is the first WiFi standard for the mmWave technology used in indoor applications. In 2018, IEEE 802.11aj was released with improved frequency bands, bandwidth (i.e., higher data rates), transmission distances, and more stable connection quality compared to IEEE 802.11ad standard. In 2021, IEEE released the 802.11ay standard\cite{mw9}, with added single-user and MIMO modes of operation in dense mmWave hotspots. The introduction of MIMO in IEEE 802.11ay offers improved performance and reliability. It is important to note that the support for up to 256-QAM high-order modulation schemes not only increases the transmission bandwidth and rate, but also improves the rate of time resolution. Furthermore, the enhanced beamforming training improves the quality and coverage of wireless signals. All these advancements have enhanced the accuracy and stability of mmWave-based positioning and sensing technology.

The positioning algorithms based on mmWave are also subject to several challenges including (i) the modeling of the channel state and accurately compensating for the received signal parameters, due to the complexity of indoor environments; (ii) the requirement for universal applicability to a variety of devices has also raised the bar for these algorithms; and (iii) detection reliability and robustness of positioning and sensing in an environment with mobility and other sources of noise for mmWave radar-based systems.

\subsection{THz}
With increasing data traffic within wireless communication networks, THz is emerging as a potential solution for providing ultra-broadband capabilities for 6G. The THz spectrum ranges from 0.1 THz to 10 THz, which fills the utilization gap between mmWaves and optical bands. THz-based PSs have attracted increasing attention for two main reasons: (i)   accurate positioning, which is a prerequisite in THz communications because of resource allocation, beamforming, and channel estimation;  and (ii) key features such as high directionality, compact antenna arrays, and large communication bandwidth that are essential in accurate PSs. Therefore, the interaction between communication and positioning plays a key role in the THz band.

In recent years, THz PSs have received considerable attention. Various methods based on  RSS \cite{lehtomaki2022distance}, CSI \cite{fan2020structured}, AoA \cite{pan2023ris}, \textit{etc.}, have been studied in the THz band. Meanwhile, there are several features associated with THz positioning: (i) RIS plays an important role in THz- positioning, since it can overcome blocking/shadowing and path losses, thereby increasing the received power level and improving PA \cite{pan2023ris,wymeersch2020radio}; and (ii) the use of learning-based positioning methods. For instance, Fan \emph{et al.} \cite{fan2020structured} proposed a structured bidirectional long short-term memory (LSTM) recurrent NN architecture to achieve a 3D indoor positioning with a mean distance error of 0.27 m.

In early 2008, the IEEE established ”Terahertz  Interest Group” (IGthz) within the 802.15 working group. This is followed by the first IEEE standard for sub-band wireless communications IEEE 802.15.3d in 2017 \cite{petrov2020ieee}, which is an amendment of the IEEE Std. 802.15.3, providing a wireless physical layer operating up to 100 Gbit/s. In view of the main objective of IEEE 802.15.3d, which is to demonstrate the feasibility of fixed point-to-point THz communication, research on THz positioning is relatively limited. In 2019, the FCC unanimously agreed to lift restrictions on frequencies above 95 GHz, thereby allocating 21.2 GHz of spectrum for unlicensed use and authorizing experimental activities in the electromagnetic spectrum up to 3 THz. This is also beneficial for research on positioning using the THz band. At THz frequencies, there are significant challenges in terms of hardware imperfections and synchronization. Furthermore, since THz signal experience substantial path losses, their design must be carefully tailored to meet the needs of users with a range of performance requirements, thereby maximizing energy efficiency. Moreover, a realistic THz channel model is still required that comprehensively addresses the THz-specific characteristics, such as LOS, NLOS and hardware impairments.

\subsection{VLP}

The RF-based PSs are less accurate mostly due to multipath induced fading and signal penetration. Optical wireless technology-based PSs utilizing infrared (IR), ultraviolet, and visible bands have been introduced in recent years with high PA. Note, at low levels, all light sources are harmless to humans and depending on the wavelength have different uses in many applications. The IR technology has been used for PSs with active beacon transmitters or receivers placed at known locations and mobile transmitters or receivers with unknown positions \cite{HTC2018}. In \cite{zhang2012microsoft}, Microsoft Kinect has used a continuously projected IR structured light to detect the environment using an infrared camera. The implementation of RSS-based IPSs is simpler compared with TOA and AoA, since (i) there is no requirement for highly accurate transceiver synchronization and for a receiver with efficient detection of the incidence angle; and (ii) have high PA due to the availability of LoS paths for most indoor environments. Several challenges must be overcome, however, including the concurrent transmission of the optical signals using multiple LED light sources may make it difficult to recover the signals using a single PD-based receiver; and transmitters and receivers are often assumed to be parallel (i.e., without tilting angle) which may reduce the PA.

In contrast,  VLPs have received significant attention over the past decade, in which LED lights are used for positioning, illumination, and data communications. It used LED lights at transmitters and photodiodes (PDs) or camera sensors as the receiver. VLP offers inherent security at the physical layer since lights emitted from the sources and reflected surfaces are maintained within a confined space, abundant license-free spectrum, immunity to RF-induced electromagnetic interference, low costs, and high PA compared with the RF-based PSs \cite{zhuang2018survey}, \cite{Matheus2019visible, yang2019relay}, \cite{chaudhary2019current}, \cite{maheepala2020light}. There are numerous applications for VLP, including location tracking, navigation, vehicular communications, shelf-label advertising in supermarkets, medical surveillance, street advertising, and robot movement control \cite{dawood2021comparative}.

VLPs are categorized based on fingerprinting, proximity, triangulation, sensor-assisted, ML, and filtering techniques. In fingerprinting, also known as scene analysis, distinct features of signals together with AoA, ToA, TDoA, and RSS are used for estimating positioning. In \cite{hann2010white}, VLP using a correlation approach to match the pre-estimated address for each LED light with the detected signals at the receiver was investigated experimentally in an indoor environment with PA of 1.495 cm. In \cite{yang2013visible}, VLP with time division multiplexing was proposed to mitigate interferences with an average PA of 1.68 cm. The proximity method is very simple but with the PA as good as the resolution of the grid and the number of transmitter reference nodes. For example, in \cite{del2013vlc} VLP based on the LED light and a mobile phone was proposed for to determine the precise location. Both passive and active beacons were investigated with error-free range of up to 4.5 m. Using LED lights and a geomagnetic sensor, in \cite{nakajima2013new} VLP was adopted to accurately determine position and travel directions for visually impaired people. Based on rotation matrix and support vector machines, the precise limits of field of view as well as azimuth and tilt angulations were calculated with 80\% less computation than conventional geometric optics \cite{sertthin2011physical}.

In triangulation, the target’s position is determined by distance measurement from at least three reference locations using RSS, TOA, TDOA, and direct detection techniques \cite{wang2013position,kim2012indoor,yang2014three,zhuang2018survey}. Perfect synchronization between the transmitter (Tx) and receiver (Rx) is required for TOA and TDOA \cite{ott1977vehicle,del2017survey}. In RSS, the optical receiver should receive signals from multiple LED transmitters with no interference. Note that the coordinates of LED transmitters in the real world are unknown prior to determining the position of the receiver. Therefore, it is critical to establish the link between the LEDs and the receiver to obtain the coordinates of the LEDs. The implementation of RSS-based IPSs is simpler compared with TOA and AoA, since (i) there is no requirement for highly accurate transceiver synchronization and for a receiver with efficient detection of the incidence angle; and (ii) have high PA due to the availability of LOS paths for most indoor environments. Several challenges must be overcome, however, including the concurrent transmission of the optical signals using multiple LED light sources may make it difficult to recover the signals using a single PD-based receiver; and transmitters and receivers are often assumed to be parallel (i.e., without tilting angle) which may reduce the PA.

Since in VLC-IPS the transmission data rate is not an issue, both camera (image sensor) and PD-based receivers could be used.

\subsubsection{PD-based VLP systems}
At the transmitter, the encoded address and identification (ID) information of each LED are broadcast via free space. At the receiver, the optical signals are detected using a PD-based optical receiver for regeneration of the electrical signal. 
The channel gain can be expressed by Lambertian model \cite{yang2019relay}.
From the perspective of measurement, PD-based VLP algorithms can be classified into several categories: i) Proximity\cite{cherntanomwong2015proximity,del2013Proximity}, ii) TOA/TDOA \cite{wang2013TOA,do2014tdoa, Du2018tdoa}, iii) AOA\cite{zhang2022beacon}, iv) RSS\cite{shen2022hybrid}, and v) Fingerprinting\cite{oh2022vlc}. 


\subsubsection{Image Sensor (IS)-based VLP systems}
Different from the PD that relies on the Lambertian channel model, IS-based VLP systems rely on capturing the images of intensity modulated the LED luminaire and using image processing algorithms to determine the required position of objects and people \cite{wang2013position}. The information on the LED light in the image is provided based on the image coordinates. A wide usage of cameras including those in smart devices can be used in IS-based VLPs. The IS-based VLPs have several unique features compared to PD-based systems, such as a larger field of view and spatial and wavelength separation of light \cite{hernandez2022wifi}. 
A complementary metal-oxide semiconductor (CMOS) camera is typically used in IS-based VLP systems. The rolling shutter exposure model of CMOS camera can help decode the VLC information by capturing black and white stripes. Additionally, the camera can also capture the visual information of the LED luminaires for analyzing the geometric relationship between the LED luminaires and the receiver. This characteristic has been taken into account by several recent works \cite{huang2022three, bai2021vp4l, wang2021arbitrarily, zhu2023doublecircle}.
For instance, Huang \textit{et al.} \cite{huang2022three} proposed to use camera to capture reflected lights of a single LED luminaire from the floor, and the highlights were regarded as the projections formed by virtual LEDs and deriving a geometric relationship between two virtual LEDs for final position estimation. 
In addition, there are also the contour shapes of the luminaire considered for VLP  when an IS-based receiver is used. Bai \textit{et al.} \cite{bai2021vp4l} considered exploiting the rectangular features of a single luminaire an IS-based VLP algorithm. The circular luminaire features were also used to estimate the orientation and location of the receiver \cite{wang2021arbitrarily, zhu2023doublecircle}. 

 A variety of fusion algorithms have emerged to exploit the advantages of the single PD- and IS-based VLP system in recent years. 
Some works focused on fusing RSS and image sensing to achieve positioning by simultaneously using PD- and IS-based receivers. For instance, Hua \textit{et al.} \cite{hua2021fusionvlp} introduced the fusion VLP system that leveraged ensemble KF to fuse the measurements from the PD and camera for real-time positioning. 
Bai \textit{et al.} \cite{bai2020received, bai2021ecarssr} proposed the use of measurements from the camera to provide incident angle information to RSSR algorithm so that the receiver can be located regardless of orientation. 
In addition, researchers have tried to fuse AOA and image sensing \cite{Cincotta2019}, in which, the incident angle derived from the image was used by the AOA algorithm. In \cite{aparicio2022experimental}, a triangulation algorithm based on AOA and RSS measurements was proposed to estimate the receiver's position by implementing the least squares estimator and trigonometric considerations.  
Overall, the fusion of different VLP measurements makes the system more accurate and practical, such as reducing the required LED luminaires, and relaxing the orientation limitation of the receiver. 


Despite centimeter-level accuracy, VLP still faces the challenges of industrialization, reliability, and cost challenges. These include: (i) impact of the transmitter tilting angles; (ii) limited frame rates of, therefore limited data rates; (iii) light flickering; (iv) multipath reflections.
In practice, VLP is susceptible to occlusion, ambient light, and other environmental factors, which may lead to positioning failure.
Moreover, integrating a VLP system necessitates merging with existing frameworks, such as building management systems or mobile applications. Table \ref{tab_vlp} shows the IS-based VLP.

\begin{table*}
\setlength{\tabcolsep}{3.5pt}
\renewcommand{\arraystretch}{2}
\centering
\caption{IS-based VLPs}
\label{tab_vlp}

\begin{threeparttable} 

\fontsize{9}{10}\selectfont
\begin{tabular}{|c|c|c|c|}

\hline
Scheme & PA(cm) & Dimension(m)  & Complexity  \\
\hline
OCC	\cite{le2019photography}& 10 & 2 & $\bigodot$\\ \hline
\makecell{RS-OCC \& multiple FSK \cite{shahjalal2018implementation} }	& 2& 2.6&● \\ \hline
\makecell{AoA \& RSS with \emph{k}-nearest neighbors in feature\\ space algorithm \cite{zheng2018high}} & 1.97&  $0.7\times0.3\times0.2$  &●  \\ \hline
\makecell{AoA \& RSS with a geomagnetic field sensor\\ and an accelerometer \cite{lee20193d} }& \textless 10&  $1\times1\times2.4$ &● \\ \hline
\makecell{LED + RS \& piecewise fitting \cite{ji2019single}}& \makecell{3.17(2D) \\4.45(3D)}	&1.2	&$\bigodot $\\ \hline
\makecell{Inertial measurement unit and IS \cite{saadi2016led} }	& 16 & $1.8\times1.8\times2$ &$\bigodot $\\
\hline 
\end{tabular}
   \begin{tablenotes}
   \centering
   \footnotesize
		\item {$^{\rm *}$We roughly categorize KPI values in the table into 3 levels, e.g. $\bigcirc$: Low, $\bigodot$: Medium, ●: High.  }
     \end{tablenotes} 
\end{threeparttable} 

\end{table*}

\subsection{Hybrid RF-optical}
Researchers are advocating the development of a hybrid PS that combines the advantages of visible light and RF signals to harvest the advantages of both.
The current mainstream fusion positioning solutions have successfully integrated VLP with RF technologies such as WiFi, 5G, and Bluetooth, as reported in the literature \cite{survey-wifi, zigbee, vlp-5g, yby2, albraheem2023hybrid, bluetooth}.

For instance, a heterogeneous PS incorporating LiFi and WiFi was conceptualized to enhance indoor PA\cite{survey-wifi}.
In addition, Shi \textit{et al.} \cite{vlp-5g} proposed a 5G IPS centered on VLC and broadband communications, specifically designed for museum applications. The system utilized unlicensed visible light to provide visitors with high-accuracy positioning on a mobile device, achieving a mean positioning error of 0.18 m. 

Combined with Bluetooth, a hybrid PS was introduced in \cite{albraheem2023hybrid}, where the initial location based on VLC proximity was collected prior to, determining the location of the receiver using Bluetooth RSS trilateration, yielding a notable accuracy of up to  0.03 m. Another approach by Luo \textit{et al.} \cite{yby2} involved a spring model based on Bluetooth signals for hybrid VLP and Bluetooth positioning. The intensity of visible light signals was detected through the Bluetooth beacon set in advance to match the fingerprint database. Simulation results showed that the system can achieve an average PA of 6 cm. 
Hussain \textit{et al.} \cite{bluetooth} used a VLC-based indoor mapping application to facilitate Bluetooth MAC address mapping. In this way, the advantages of VLC and Bluetooth can be combined to achieve superior positioning performance. The key features of the existing positioning technologies are summarized in Table \ref{tab3}.

\begin{table*}[htbp]
\renewcommand{\arraystretch}{2}
\setlength{\tabcolsep}{1.8pt}
\scriptsize
\centering
\caption{Positioning Technology Comparison}
\label{tab3}
\fontsize{7.5}{10}\selectfont
\begin{threeparttable} 

\begin{tabular}{|c|c|c|c|c|c|c|c|c|c|c|c|}

\hline
\multirow{2}{*}{Technology}  & \multicolumn{10}{c|}{Key Performance Indicators (KPI)} & \multirow{2}{*}{Comments}\\ 
\cline{2-11} 
 & PA (m) & Coverage & \makecell{Power\\Consumption} & Cost & Real-time & Computation & Availability & Robustness & \makecell{Security and\\Privacy} & Complexity &\\
\hline

\makecell{Celluar\\Networks\cite{shao2018indoor}}  & 1-2  & ● & $\bigodot$  & $\bigodot$ & Soft & $\bigodot$ & ● &  $\bigodot$&  $\bigodot$& ● &\makecell[l]{Very High Coverage,\\relatively high accuracy\\with low power consumption. } \\ \hline

\makecell{WiFi \cite{qin2021ccpos}} &\makecell{1-5}  & $\bigodot$ &●  & $\bigcirc$ & Hard & $\bigodot$ & ● & $\bigcirc$&  $\bigodot$&  $\bigodot$&\makecell[l]{Environment dependant, \\large database, limited\\ coverage range and mobility. } \\ \hline

Bluetooth \cite{wang2013bluetooth}  &\makecell{1-5}  & ● &$\bigcirc$  & $\bigcirc$ & Hard/Soft & $\bigodot$ & ● & $\bigcirc$&  $\bigodot$&$\bigodot$ &\makecell[l]{High coverage, low power,\\but is unstable and easily\\affected by radio interference. }  \\ \hline

RFID \cite{shen2016survey} &\makecell{1-2} & $\bigodot$ &$\bigcirc$  & $\bigcirc$ & Soft & ● & $\bigodot$ & $\bigcirc$& $\bigcirc$&  $\bigodot$ &\makecell[l]{Low power consumption,\\ limited mobility, but low\\ security and high delay.}\\ \hline

UWB\cite{kuhn2010adaptive}  &\makecell{0.1-1}& $\bigodot$ &$\bigodot$  & ● & Hard &  $\bigodot$ & $\bigcirc$ & ●& ●& $\bigodot$ &\makecell[l]{High costs, \\limited coverage range. }\\ \hline

mmWave\cite{mw5} &\makecell{0.1-10}& $\bigodot$ &$\bigodot$  & $\bigodot$ & Soft/Hard &  $\bigodot$ & $\bigodot$ & $\bigcirc$& $\bigodot$&$\bigodot$ &\makecell[l]{Also widely used in radar-\\based sensing, particularly,\\ besides positioning. }\\ \hline

THz \cite{fan2020structured} &\makecell{0.1-10}& $\bigcirc$ &$\bigodot$  & ● & Hard &  $\bigodot$ & $\bigcirc$ & $\bigcirc$& $\bigodot$&$\bigodot$ &\makecell[l]{Faces hardware and \\synchronization issues, and is \\still under experiment. }\\ \hline

VLP \cite{hann2010white, del2013vlc} &\makecell{\textless 0.05}& $\bigcirc$ &$\bigodot$  & $\bigodot$ & Soft &  $\bigodot$ & $\bigodot$ & $\bigcirc$& $\bigodot$&  $\bigcirc$&\makecell[l]{Modifying existing LED \\light, blocking. }\\ \hline

\makecell{Hybrid\\RF-optical \cite{vlp-5g}} &\makecell{0.01-1}& $\bigcirc$ &$\bigodot$  & $\bigodot$ & Soft &  ● & $\bigodot$ & $\bigodot$&$\bigodot$& ● &\makecell[l]{Hybrid systems based on VLP\\and RF-based systems. }\\ \hline

\end{tabular}
    \begin{tablenotes}
         \footnotesize
         \centering
		\item {$^{\rm *}$We roughly categorize KPI values in the table into 3 levels, e.g. $\bigcirc$: Low, $\bigodot$: Medium, ●: High.  }
    \end{tablenotes} 
\end{threeparttable} 
\end{table*}

\section{Challenges}
This section summarizes the key challenges of current PSs.
The challenges and pitfalls of PSs require technological innovation and interdisciplinary integration to improve the link reliability and achieve PA, which are outlined in the following.

\subsection{Trade Off Between Accuracy and Cost}
PA and cost are essential factors in the design of PSs, yet achieving a satisfactory balance between these two remains a challenge. 
For instance, systems such as cellular networks and WiFi offer the advantage of low-cost positioning by leveraging the existing infrastructure. However, their accuracy may not meet the demands of applications like AR/VR that require centimeter-level accuracy. Conversely, systems like UWB and VLP can achieve accurate positioning. Nonetheless, they necessitate additional infrastructure, leading to higher costs. In particular, the high deployment cost is a prominent issue in UWB \cite{li2022optimal}. As for VLP, while the cost of retrofitting each light is insignificant, due to the limited coverage range of each light, the cost of large-scale deployment still needs further verification. Therefore, the quest to reconcile the trade-off between accuracy and cost continues to be a daunting task in the realm of indoor positioning. This trade-off is crucial for broadening the applicability of positioning technologies across various sectors. 

As the field progresses, it is essential to focus not only on developing new algorithms but also on enhancing the cost-effectiveness of the system. It is through this dual approach that technological advancements will be both practically applicable and economically viable, thus enabling broader implementation and accessibility. By prioritizing the development of cost-effective solutions along with cutting-edge algorithmic improvements, it is possible to drive the widespread adoption of IPSs. As a result of this strategy, high-precision technologies will become more accessible to a broad range of applications, ranging from consumer electronics to industrial automation, thereby bridging the gap between theoretical excellence and practical applications.

\subsection{Trade Off Between Coverage and Accuracy}
Positioning environments are often characterized by their complexity, especially in indoor scenarios. Moreover, these environments are often cluttered with obstacles such as walls, furniture, and human movement, which can obstruct signals and cause issues such as multipath propagation. This complexity means systems with large coverage, such as cellular networks, tend to suffer from limited accuracy due to the long propagation path between the transmitter and the receiver. In contrast, systems such as THz and VLP, which offer limited coverage are reported to achieve centimeter-level PA. 
Note that short propagation paths ensure simple transmission links but also limit their availability and robustness, making them less versatile in various scenarios. 
Therefore, the trade-off between the coverage and the accuracy limit the applicability of existing PSs. 

To navigate this trade-off, future developments should focus on innovative approaches that can either extend the effective coverage of high-accuracy PSs or enhance the PA of wide-coverage PSs. 
For instance, by combining multiple positioning technologies, it may be possible to leverage their respective strengths in order to achieve promising pathways. In addition, it is also a possible way to employ advanced signal processing and ML algorithms to mitigate the effects of signal obstruction and multipath propagation. The next generation of PSs can achieve wide coverage and high accuracy by pushing the boundaries in these areas, thus enhancing their utility across a broader range of applications.

\subsection{Security and Privacy}
Human and device location information is considered as sensitive data that can expose users to a variety of risks including stalking, theft, and even security threats. Location security and privacy are essential components of comprehensive cybersecurity efforts.
These efforts are dedicated to safeguarding the confidentiality, integrity, and availability of geographical information, which is becoming increasingly pivotal in the development of new applications.
However, security and privacy issues in positioning have not garnered as much focus as those in the field of communications. 
Since PSs often operate within strict energy constraints, they are unable to employ complex methods for ensuring the privacy and security of location data. 
Moreover, PSs may use diverse technologies based on different methodologies, and each of them has its own vulnerabilities and security implications. This diversity complicates the tasks of creating a universal solution for security and privacy. 

From a technological perspective, enhancing location data security requires a multi-faceted approach. This could involve the development of lightweight cryptographic algorithms suitable for energy-constrained devices, advanced anonymization techniques to protect user identities, and robust access control mechanisms. Additionally, standardized security protocols across different positioning technologies should also be considered to ensure a cohesive and secure framework. By addressing these challenges, it is possible to foster trust and promote broader adoption of indoor positioning applications, balancing the benefits of precise location services with the imperative of protecting individual privacy and security.

\subsection{Complex and Dynamic Environments}
Positioning environments change over time. ML-based methods have been applied to dynamically update parameters based on the data for continuous improvement and adaptation to environmental changes. In addition, ML-based methods are used to effectively integrate and process data from various sources. These methods, however, typically require a large amount of labeled data, which is closely related to the environment and can be labor-intensive. Complex and dynamic environments can adversely affect the performance degradation. On one hand, the controlled environment in existing methods can differ from the practical environment. On the other hand, long-term changes in the environment may lead to inaccurate tag data, thereby affecting the results of position estimation. Therefore, positioning methods need to adapt to variable and complex environments and reduce the reliance on labels. 

To overcome these challenges, semi-supervised or unsupervised learning can be used to learn from limited or unlabeled data. In addition, adaptive models are expected to be developed for PSs that can dynamically update their parameters in response to environmental changes, to enhance their effectiveness in the face of the variability and complexity of real-world environments. With their powerful ability to understand and predict environments, large models may play a crucial role in solving these challenges.

\subsection{Diverse Requirements and Applications}
The PSs should be able to cater to a wide array of applications including those for public utilities, enterprises, and individuals, as well as applications for online and offline use, each having its own set of requirements for accuracy, latency, and scalability. There is a significant challenge in tailoring PSs to meet these diverse needs without compromising performance. Therefore, it is necessary to develop flexible positioning techniques that can be tailored to meet the needs of different users and applications. The integration of multiple data sources and sensors, for example, could enhance the ability to sense the environment, so as to meet specific accuracy, latency, and scalability requirements of different applications.


\section{Conclusion}
In this paper, we provided a comprehensive review of existing positioning technologies. To begin with, we reviewed the evolution of positioning over wireless networks. Then, we discussed the applications of positioning technology from the perspectives of public facilities, enterprises, and individuals. Next, we have summarized the existing KPIs and measurements for positioning and conducted a detailed comparison. We further investigated the key techniques of positioning such as large models, adaptive systems, and RIS, which may significantly enhance the performance of a PS in the future. As a step forward, we discussed various typical wireless positioning technologies. We not only focused on the progress of these technologies in the academic community but also covered their standardization process. Meanwhile, we provided an in-depth comparison of these technologies and summarized the KPIs that each technology needs to focus on more. Finally, we summarized the key challenges of positioning systems. Although positioning technology currently still faces many challenges, we firmly believe that positioning will play an increasingly important role in wireless networks in the future.

\bibliographystyle{IEEEtran}
\bibliography{main}

\end{CJK}
\end{document}